\documentclass[prb,aps,amssymb,showpacs,twocolumn,hyperref]{revtex4-1}
\usepackage{color}
\usepackage{graphicx}
\usepackage{dcolumn}
\usepackage{bm}
\usepackage{hyperref}
\usepackage{amsmath}
\usepackage{latexsym}
\usepackage{amssymb}
\usepackage{layout}
\usepackage{verbatim}
\usepackage{amsfonts,epsfig}
\begin{document}
\title{Nuclear spin polarization in a single quantum dot pumped by two laser beams}
\date{\today}
\author{Xiao-Feng Shi}
\email[E-mail: ]{xshi@physics.ucsd.edu}
\affiliation{Department of Physics, University of California San Diego, La Jolla, California 92093-0319, USA}

\begin{abstract}
We theoretically investigate dynamic nuclear spin polarization in a self-assembled quantum dot pumped optically by two laser beams. With the assumption that a noncollinear interaction between the hole spin and nuclear spins leads to nuclear spin polarization, we find that both weak and strong nuclear spin polarizations can arise, depending on the intensities and central frequencies of the lasers. For weak nuclear spin polarization, we use a perturbation method to show that the distribution of the nuclear spin Overhauser field may become significantly narrower. Using Monte Carlo simulations to study a single quantum dot, we find that strong nuclear spin polarization can also be generated via appropriate optical pumping.
\end{abstract}
\maketitle

\section{introduction}
Dynamic nuclear spin polarization by optical pumping on single quantum dots~(QDs) is extensively studied in recent years,\cite{OpticalOri,Gammon2001,Bracker2005,Braun2006,PhysRevLett.96.167403,Korenev2007,Greilich2007,Tartakovskii2007,Maletinsky2007,PhysRevB.74.081306,Skiba-Szymanska2008,PhysRevLett.100.056603,Xu2009,Latta2009,Nikolaenko2009,Huang2010,Makhonin2010,Issler2010,Chekhovich2010prb,Chekhovich2010,Chekhovich2011,Barnes2011,Makhonin2011,Korenevemail,Sun2012,Hogele2012,LeGall2012,Glazov2012,Kaji2012,Urbaszek2013} due to its importance for improving techniques in quantum computing, and for understanding the nuclear spin environment of the electron~(hole) spin qubit. For instance, an efficient suppression of nuclear spin fluctuation by optical pumping allows one to prolong the electron spin coherence time in a QD.\cite{Xu2009,Sun2012} Also, significant nuclear spin polarization may be achieved\cite{Gammon2001,Bracker2005,Skiba-Szymanska2008} even by optical excitation of spin-forbidden transitions.\cite{Chekhovich2010} In order to understand the dynamic nuclear spin polarization induced by continuous optical pumping on single QDs, a microscopic theory\cite{Yang2012,WenPRB} was recently introduced to explain relevant experiments.\cite{Latta2009,Hogele2012} However, this theory is restricted to single laser pumping and only treats weak nuclear spin polarization, while experiments involving two laser beams\cite{Xu2009,Sun2012} or strong nuclear spin polarization remain to be explained microscopically.

In this paper, we study dynamic nuclear spin polarization in a single QD charged with one electron and pumped by two narrow-linewidth continuous wave lasers~(see Fig.~\ref{system01}). We find that both weak and strong nuclear spin polarizations can be generated. Specifically, when the laser with central frequency $\omega_{1(2)}$ and moderate Rabi frequency $\Omega_{1(2)}$ is off-resonant~(resonant) with the transition between the exciton and the electron spin eigenstate $|x-(x+)\rangle$, the nuclear spins tend to have weak polarization for $\Omega_{1}\ll\Omega_{2}$, but strong polarization for $\Omega_{1}\gg\Omega_{2}$. In the former case, we derive a Fokker-Planck equation\cite{Carmichael1999} for the evolution of the probability density of nuclear spin polarization. We use the Fokker-Planck equation to show that the nuclear spin fluctuation can be reduced, thus enhancing the electron spin coherence. Also, we perform numerical study of several interesting phenomena observed in laser spectroscopy experiments.\cite{Xu2009} For the case of strong nuclear spin polarization, the perturbation method for deriving the Fokker-Planck equation breaks down. We then create a toy model comprising a small QD in order to obtain unbiased results using the Green's function Monte Carlo simulation.\cite{Hetherington1984,PhysRevB.57.11446} Our Monte Carlo simulation shows that at least as high as $50\%$ of the nuclear spin polarization degree can be generated by pumping optically on the single QD in our toy model. We note that a $50\%$ degree of nuclear spin polarization is close to the large nuclear spin polarization degree observed in experiments by optically pumping single QDs.\cite{Gammon2001,Urbaszek2007,Skiba-Szymanska2008,Chekhovich2010}

The paper is organized as follows. Section II briefly introduces the system Hamiltonian. Section III solves the equation of motion for the nuclear spin population. In Sec.~IV, we study the case corresponding to the experiments in Ref.~\onlinecite{Xu2009}, where only weak nuclear spin polarization was observed. In Sec.~V, we first identify the factors responsible for large nuclear spin polarization. Then we numerically study a small QD of artificial size to show that large nuclear spin polarization can indeed be generated by optical pumping on single QDs. We conclude in Sec. VI.

\section{Model}
\begin{figure}
  \centering
  \includegraphics[width=2.5in]{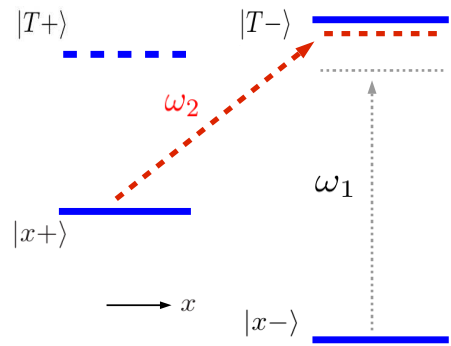}
  \caption{(Color online) Schematic of an electron-hole system. The $x$-polarized laser connecting $|x-\rangle$ and $|T-\rangle$ has central frequency $\omega_1$ and Rabi frequency $\Omega_1$, and the $y$-polarized laser connecting $|x+\rangle$ and $|T-\rangle$ has central frequency $\omega_2$ and Rabi frequency $\Omega_2$.}
  \label{system01}
\end{figure}

Consider a self-assembled InAs QD charged with one electron, where the confined electron interacts with $N$ indium nuclear spins and $N$ arsenic nuclear spins. Defining $z$ as the growth direction of the QD, we apply an in-plane static magnetic field $\mathbf{B}$ in the $x$ direction and label the electron spin eigenstates as $|x\pm\rangle$, as shown in Fig.~\ref{system01}. Two linearly polarized coherent laser beams, one with the polarization of the electric vector along $x$ direction, and the other along the $y$ direction, selectively couple the two electron spin states to a common trion level, denoted as $|T-\rangle$. The trion consists one heavy hole and two electrons,\cite{Winkler2003} with the two electrons in the singlet configuration. The Hamiltonian of the system is
\begin{eqnarray}
  \hat{H}&=&\hat{H}_{\text{eh}}+ \hat{H}_{\text{n}}+ \hat{H}_{\text{\tiny HI}},\label{e01}
\end{eqnarray}
where $\hat{H}_{\text{eh}}$ is the Hamiltonian of the electron-hole system, $\hat{H}_\text{n}$ the nuclear spin Zeeman term, and $\hat{H}_{\text{\tiny HI}}$ the hyperfine interaction between the electron/hole spin and the nuclear spins in the QD,\cite{Slichter1992,Eble2009,Leger2007}
\begin{eqnarray}
  \hat{H}_\text{\tiny HI}&=&
  \sum_ja_{\text{e},j}\left(\hat{S}_{\text{e}}^x\hat{I}_j^x+ \hat{S}_{\text{e}}^y\hat{I}_j^y+ \hat{S}_{\text{e}}^z\hat{I}_j^z \right)\nonumber\\
&&+ \sum_j \frac{a_{\text{h},j}  }{1+ |\beta|^2} \left\{\hat{S}_{\text{h}}^z\hat{I}_j^z+\frac{2|\beta|}{\sqrt{3}}\left[\hat{S}_{\text{h}}^x(\hat{I}_j^x\cos\delta \right.\right.
\nonumber\\&&\left.\left.+\hat{I}_j^y\sin\delta)+\hat{S}_{\text{h}}^y(\hat{I}_j^y\cos\delta -\hat{I}_j^x\sin\delta)   \right]  \right\},\label{e02}
\end{eqnarray}
where $a_{\text{e(h)},j}$ is the hyperfine interaction strength between the electron~(hole) spin $\mathbf{S}_{\text{e(h)}}$ and the nuclear spin $\mathbf{I}_j$, the superscript $x,y$ or $z$ denotes component of the spin moment in the corresponding direction, and $\beta=|\beta|e^{i\delta}$ is the heavy-light hole mixing coefficient.\cite{PeterYYu2001, Koudinov2004,Leger2007,Eble2009} Besides the dipole-dipole coupling in the hole spin-nuclear spin interaction, there are extra hyperfine couplings with strength proportional to $|\beta|$ in Eq.~(\ref{e02}) that arise from the mixing between the heavy and light hole bands. The band-mixing is caused by an in-plane strain of the QD when one grows QDs in the Stranski-Krastanow growth mode during molecular beam epitaxy.\cite{Testelin2009,Leger2007} Observed values of $|\beta|$ for typical In(Ga)As QDs range from $0.02$ to $0.7$.\cite{Xu2009,Eble2009, PhysRevB.72.161312,PhysRevB.77.075317} The angle $\delta$ is determined by the strain detail\cite{Testelin2009,Leger2007} of the QD and can in principle take any value between $0$ and $2\pi$. For simplicity, we have ignored the intrinsic interactions~(including dipole-dipole interactions) between nuclear spins in $\hat{H}_{\text{n}}$. The latter will be considered later when studying relevant experiments, involving nuclear spin depolarization.\cite{Gong2011}

For study involving small nuclear spin polarizations, we consider a QD containing $N=9500$ InAs molecules~(estimated from Ref.~\onlinecite{Xu2009}). Without loss of generality, we assume $g_{\text{e(h)}}=0.49~(-0.13)$. Although this choice of parameters is made in order to study the experiments in Ref.~\onlinecite{Xu2009}, we note that our general results for both small and large nuclear spin polarizations do not rely on these specific assumptions. In the electron-hole system shown schematically in Fig.~\ref{system01}, the laser connecting $|x-\rangle$ and $|T-\rangle$ may be slightly off-resonant with the transition between the electronic and excitonic states, while the laser connecting $|x+\rangle$ and $|T-\rangle$ is resonant unless otherwise specified. For this case with a strong external magnetic field of $|B|=2.64$~T, we can prove that the level $|T+\rangle$ decouples from the rest of the system~(Appendix~\ref{steadystate}), resulting in a three-level electron-hole system which we call a $\Lambda$ system~($\Lambda$S).

\section{Nuclear spin dynamics }
Following Refs.~\onlinecite{Xu2009,Latta2011,Hogele2012}, we assume that the effective noncollinear hyperfine interactions between the electron/hole spin and nuclear spins can lead to dynamic nuclear spin polarization in a single QD. For simplicity, we only consider the hole spin-nuclear spin noncollinear interaction of the form $\hat{S}_{\text{h}}^x\hat{I}_j^y$ here, which appeared in Eq.~(\ref{e02}). We neglect the influence of the transverse part of the electron~(hole) spin-nuclear spin hyperfine interaction on the nuclear spin dynamics as discussed in Appendix~\ref{steadystate}.

Using the theory developed in Ref.~\onlinecite{Yang2012}, we derive an equation for the time evolution of the nuclear spin population $\hat{P}(t)$, i.e., the diagonal part of the reduced density matrix of the nuclear spins,
\begin{eqnarray}
  \frac{d}{dt}\hat{P}(t)   &\approx &  -\sum_{j}  \left\{  \left[   \hat{I}_j^- ,~~ \hat{I}_j^+ \hat{W}_{\alpha_j,+}  \hat{P}(t)   \right]\right.\nonumber\\&&\left.+\left[   \hat{I}_j^+  ,~~ \hat{I}_j^- \hat{W}_{\alpha_j,-}   \hat{P}(t)   \right]\right\}
 ,\label{e09}
\end{eqnarray}
where $\alpha_j$ denotes In~(As) when there is an indium~(arsenic) atom at site $j$, and
\begin{eqnarray} 
 \hat{W}_{\alpha_j,\pm}\! &\!=\!&\!\left|\frac{|\beta| a_{\text{h},j} \sin\delta}{\sqrt{3}(1+ |\beta|^2)} \right|^2
  \int  e^{\pm ig_j\mu_{\text{\tiny N}}B t'}\nonumber\\
  &&\times \text{Tr}[ \hat{S}_{\text{h,I}}^{x}(t')\hat{S}_{\text{h}}^x \hat{\rho}_{\Lambda}^{(ss)}(0)]   dt',\label{EQ04}\\
  \hat{S}_{\text{h,I}}^x(t)&=&\hat{U}^{\dag}(t) \hat{S}_{\text{h}}^x \hat{U}(t),~\hat{U}(t)=\mathcal{T}e^{-i\int_0^t[\hat{H}_{\text{eh,r}}(t')+\hat{H}_{\text{n}}]dt'},
\nonumber
\end{eqnarray}
where $\mu_{\text{\tiny N}}$ is the nuclear magneton, $\hat{\rho}_{\Lambda}^{(ss)}$ is the reduced density matrix for the steady state of the $\Lambda$S,\cite{Yang2012} Tr denotes trace over the $\Lambda$-system degrees of freedom, $\mathcal{T}$ is the time ordering operator, $\hat{H}_{\text{eh,r}}(t')$ is obtained by performing a rotating frame transformation on $\hat{H}_{\text{eh}}$ (see Appendix \ref{steadystate}), and the integration is from $-t$ to $t$. However, since the scale of $t$ is much larger than the time scale of the $\Lambda$S, we can perform the integration from $-\infty$ to $\infty$ and the final integration can be calculated through the quantum regression theorem.\cite{Onsager1931,Lax1963}

There is a feedback loop between the nuclear spins and the $\Lambda$S. First, the rate $\hat{W}_{\alpha_j,\pm}$ in Eq.~(\ref{e09}) for flipping the nuclear spins depends on the steady state of the $\Lambda$S. Second, the steady state of the $\Lambda$S depends on the nuclear spin state, since the eigenenergies of the two electron spin eigenstates $|x\pm\rangle$ in the $\Lambda$S are shifted by an effective magnetic field contributed by $\hat{S}_{\text{e}}^x\sum a_{\text{e},j}\hat{I}_j^x$ in $\hat{H}_{\text{\tiny HI}}$. Here we neglect the nuclear spin Overhauser field contributed from the hole spin-nuclear spin hyperfine interaction since $|a_{\text{h},j}|$ is much smaller\cite{Fischer2008,Testelin2009,Chekhovich2011,Fallahi2010} than $|a_{\text{e},j}|$. The effective magnetic field, i.e., the nuclear spin Overhauser field, can be written as~[see Eq.~(\ref{eq07})]
\begin{eqnarray}
  h&=&[ \mathcal{A}_{\text{In}}Is_{\text{\tiny In}}+ \mathcal{A}_{\text{As}}Js_{\text{\tiny As}}],\label{EQ05}
\end{eqnarray}
where $I=9/2,~J=3/2$, $\mathcal{A}_{\text{In(As)}}$ is the hyperfine constant of the indium~(arsenic) nuclear spins,\cite{Liu2007} and $s_{\text{\tiny In}},~s_{\text{\tiny As}}$ are the expectation values of 
\begin{eqnarray}
\hat{s}_{\text{\tiny In}}&=&\frac{1}{NI}\sum_{j\in \text{In}} \hat{I}_j^x,~\hat{s}_{\text{\tiny As}}=\frac{1}{NJ}\sum_{j\in \text{As}} \hat{J}_j^x.\label{EQ06}
\end{eqnarray}
Because of the nuclear spin Overhauser field $h$, the nominal detunings
\begin{eqnarray}
  \Delta_{1,0}&=&E_{\overline{T}}-E_{\overline{x}}-\omega_1,~~\Delta_{2,0}=E_{\overline{T}}-E_x-\omega_2
\end{eqnarray}
are replaced by the actual detunings 
\begin{eqnarray}
  \Delta_1&=&\Delta_{1,0}+\frac{h}{2},~~\Delta_2=\Delta_{2,0}-\frac{h}{2}
\end{eqnarray}
in the optical Bloch equation~[see Eq.~(\ref{eA6})]. Here $E_{x},E_{\overline{x}}$ and $E_{\overline{T}}$ are the respective eigenenergies~(when there is no net nuclear spin Overhauser field) of $|x\rangle,|x-\rangle$ and $|T-\rangle$, and $\omega_{1(2)}$ is the central frequency of the laser beam that couples $|x\mp\rangle$ and $|T-\rangle$. For different adiabatic nuclear spin Overhauser fields on the $\Lambda$S, the Zeeman energy of the electron spin is effectively shifted by different amounts, giving different flipping rates in Eq.~(\ref{e09}). When $h$ keeps on changing, the flipping rates in Eq.~(\ref{e09}) keep on changing. This feedback loop between the $\Lambda$S and the nuclear spins in the QD will not cease whenever the distribution of $h$ is changing.

\section{Small nuclear spin polarization}
\subsection{Approximation when $|s_{\text{\tiny In(As)}}|\ll1$}
\label{IVA}
In practice, it is not easy to solve the distribution of $h$ from Eq.~(\ref{e09}) since the number of nuclear spins is large. Nonetheless, when the polarizations of the two species of nuclear spins are much smaller than $1$, we can obtain solution for the evolution of the joint probability density of the nuclear spin polarizations
\begin{eqnarray}
  p_{\text{s}}(s_{\text{\tiny In}} ,s_{\text{\tiny As}})&= &
  \text{Tr}\left[\hat{P}(t)\delta_{\hat{s}_{\text{In}},s_{\text{\tiny In}} }\delta_{\hat{s}_{\text{As}},s_{\text{\tiny As}}} \right],\label{e15}
\end{eqnarray}
where $\delta_{\hat{s}_{\text{In(As)}},s_{\text{\tiny In(As)}} }$ is the delta function. In particular, we derive a Fokker-Planck equation,\cite{Hannes1984,Carmichael1999}
\begin{eqnarray}
\frac{\partial}{\partial t}p_{\text{s}}&\approx& \sum_{\alpha=\text{In,As}}\frac{\partial   }{\partial s_{\alpha} } \left[ \frac{\partial   }{\partial s_{\alpha}} D_{\alpha} p_{\text{s}}(s_{\text{\tiny In}} ,s_{\text{\tiny As}}) -v_{\alpha}p_{\text{s}}(s_{\text{\tiny In}} ,s_{\text{\tiny As}})  \right],\nonumber\\\label{EQ09}
\end{eqnarray} 
where $v_{\alpha}$ and $D_{\alpha}$, given by Eq.~(\ref{EQB11}), are known as the drift and diffusion coefficients,\cite{Carmichael1999} respectively. From the relation between $s_{\text{\tiny In}} ,s_{\text{\tiny As}}$ and $h$ in Eq.~(\ref{EQ05}), we can obtain the distribution of $h$ from $p_{\text{s}}(s_{\text{\tiny In}} ,s_{\text{\tiny As}})$.

\begin{figure}
  \centering
  \includegraphics[width=3in]{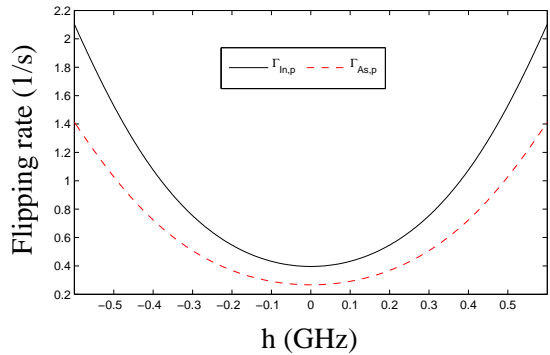}
  \caption{(Color online) Flipping rate $\Gamma_{\text{In(As),p}}$ as a function of the nuclear spin Overhauser field $h$ in a QD with $N=9500$ InAs molecules. Parameters: $B=-2.64$~T, $g_{\text{e(h)}}=0.49(-0.13),|\beta|=0.2,\delta=\pi/9, a_{\text{e},j}=-10a_{\text{h},j}=\mathcal{A}_{j}/N$, and $\gamma_{\text{s}}=0.06, \Delta_{1,0}=\Delta_{2,0}=0, \Gamma=0.4, \Omega_1=0.24,\Omega_2=1.35$~(unit: GHz). Here the gyromagnetic ratio is $g_{j}\mu_{\text{\tiny N}}=0.0093(0.0073)$~GHz$/$T, and the hyperfine constant $\mathcal{A}_{j}=13.6~(11.1)$~GHz (taken from Ref.~\onlinecite{Liu2007}) when there is an indium~(arsenic) nuclear spin at $j$.}
  \label{rate01}
\end{figure}

In order to study the absorption coefficient of the probe laser in laser spectroscopy experiments on single QDs,\cite{Xu2009} we consider the dynamics of the mean of the nuclear spin Overhauser field. To simplify the calculation, we neglect the finite width of the distribution of the polarization $s_{\text{\tiny In}}$ and $s_{\text{\tiny As}}$. Then the mean of nuclear spin polarization $s_{\text{\tiny In(As)}}$ obeys~(Appendix~\ref{AppendixFP})
\begin{eqnarray}
 \frac{d}{d t} s_{\text{\tiny In(As)}} 
&\approx&  v_{\text{\tiny In(As)} }-\gamma_{\text{dep}} s_{\text{\tiny In(As)}}  ,\label{EQ11}
\end{eqnarray}
where
\begin{eqnarray}
 v_{\text{\tiny In}} &=& -\Gamma_{\text{In},\text{p}}\left[s_{\text{\tiny In}}  -\frac{2( I+1)}{3}s_{\text{\tiny In},0}\right],\nonumber\\
\Gamma_{\text{In},\text{p}}&=& \left\langle \hat{W}_{\text{In},+}+  \hat{W}_{\text{In},-}\right\rangle ,
  \nonumber\\
 s_{\text{\tiny In},0} &=&\left\langle  \frac{\hat{W}_{\text{In},+}-  \hat{W}_{\text{In},-} }{\hat{W}_{\text{In},+}+  \hat{W}_{\text{In},-} }\right\rangle, \label{e012301}
\end{eqnarray}
and similarly for $ v_{\text{\tiny As} }$. Here, $\langle\cdots\rangle$ denotes the respective expectation value. In Eq.~(\ref{EQ11}) we have added a nuclear spin depolarization channel with rate $\gamma_{\text{dep}}\geq0$. The reason for adding this term is that, in the nuclear spin Hamiltonian $\hat{H}_{\text{n}}$, we have ignored the direct dipole-dipole interactions which play an important role in depolarizing the nuclear spins.\cite{Gong2011} A similar depolarization term was also used in the numerical fitting of the experimental results in Ref.~\onlinecite{Xu2009}. By choosing the electron spin dephasing rate $\gamma_{\text{s}}=0.06$~GHz~(this value is taken from the supplementary material of Ref.~\onlinecite{Xu2009}), the rate of spontaneous decay from $|T-\rangle$ to either $|x+\rangle$ or $|x-\rangle$ as $\Gamma=0.4$~GHz, the Rabi frequency as $\Omega_{1(2)}=0.24~(1.35)$~GHz for the probe~(pump) laser beam, we calculate the flipping rate $\Gamma_{\text{In(As)},\text{p}}$ as a function of $h$ for $\Delta_{1,0}=\Delta_{2,0}=0$, and obtain the results shown in Fig.~\ref{rate01}.

\begin{figure}
  \centering
  \includegraphics[width=3.4in]{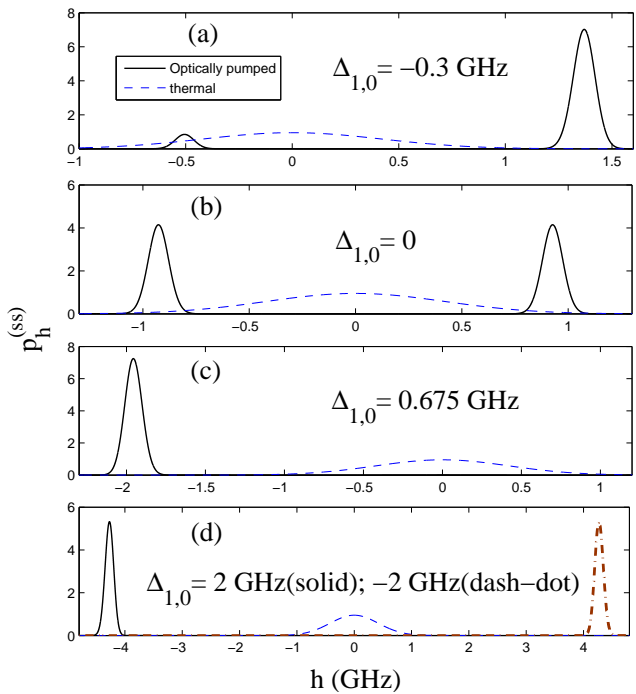}
  \caption{(Color online) Solid~(dashed) curves show the steady distributions of the nuclear spin Overhauser field $h$ built up with~(without) optical pumping. $\Delta_{1,0}=-0.3,0,0.675$ and $\pm2$~GHz in (a), (b), (c) and (d), respectively. Other parameters are the same with Fig.~\ref{rate01}, except $\mathcal{A}_{\text{In}}=14.2$~GHz here.
}
  \label{fig2_3}
\end{figure}

\subsection{Numerical results}
There are two important observations made in Ref.~\onlinecite{Xu2009}. The first is an enhanced electron spin coherence time, which was observed indirectly via a deeper dark state dip of the probe laser absorption. The second observation concerns a series of phenomena caused by dynamic nuclear spin polarization, which were observed via the change of the probe laser absorption in reference to the expected absorption when there is no dynamic nuclear spin polarization. We will study these two types of phenomena.

For the first observation where optical pumping reduces the nuclear spin fluctuation, we will study the steady distribution of the nuclear spin Overhauser field.

For the second observation, we will simulate the time evolution of the absorption coefficient of the probe laser as the laser frequency is tuned. The input for the simulation is the laser detuning, which partially determines the steady state of the electron-hole system. The output is the expectation value of the absorption coefficient for the probe laser. Since the absorption of the probe laser is proportional to the imaginary part of the density matrix element $\langle x-|\hat{\rho}_{\Lambda}^{(ss)}| T-\rangle$, we define\cite{Scully97,Berman2011}
\begin{eqnarray}
\chi_i &=&\text{Im} |\langle x-|\hat{\rho}_{\Lambda}^{(ss)}| T-\rangle|,
\end{eqnarray}
as the absorption coefficient of the probe laser. The key point is to calculate the expectation value of the nuclear spin Overhauser field, since the actual detunings for the optical transitions take into account the nuclear spin Overhauser field, which shifts the electron spin Zeeman energy. As for the input parameters, $\gamma_{\text{s}}$ is in principle determined by the fluctuation of $h$. However, as mentioned in Sec.~\ref{IVA}, we neglect the finite width of $p_{\text{s}}(s_{\text{\tiny In}} ,s_{\text{\tiny As}})$ when studying the second type of observations. This means that we have lost the information on the evolution of the nuclear spin fluctuation. As a result, we apply a constant electron spin decoherence rate $\gamma_{\text{s}}$ taken from the estimation in Ref.~\onlinecite{Xu2009} for each set of chosen parameters.

\subsubsection{Narrowed distribution of the nuclear spin field}\label{narrowedSec}
Following Ref.~\onlinecite{Yang2012}, we assume that the system contains only indium nuclear spins, since the momentum of an indium nuclear spin is three times that of an arsenic nuclear spin, and $\mathcal{A}_{\text{In}}>\mathcal{A}_{\text{As}}$. In this case, we rescale the indium nuclear spin hyperfine constant as shown in Appendix~\ref{AppInhomo}. Equation~(\ref{EQ09}) becomes
\begin{eqnarray}
 \frac{\partial}{\partial t}p_{\text{s}}&=& \frac{\partial   }{\partial s_{\text{\tiny In}} } \left[ \frac{\partial   }{\partial s_{\text{\tiny In}} } D_{ \text{In}} p_{\text{s}} -v_{\text{\tiny In}}p_{\text{s}} \right] , \label{e014}
\end{eqnarray}
where the difussion coefficient is given by 
\begin{eqnarray}
 D_{\text{In}} &=& \frac{\Gamma_{\text{In,p}}}{2NI}\left[\frac{2( I+1)}{3}- s_{\text{\tiny In}}s_{\text{\tiny In,0}}\right].\label{EQ15}
\end{eqnarray}
As in Refs.~\onlinecite{Hannes1984, Yang2012, Rudner2007}, the solution for the steady distribution of nuclear spin polarization is
\begin{eqnarray}
p_{\text{s}}^{(\text{ss})}(s_{\text{\tiny In}} )&=&\frac{\mathcal{N}}{D_{\text{\tiny In}}(s_{\text{\tiny In}})} \text{exp}\left[ \int_{-1}^{s_{\text{\tiny In}} } \frac{v_{\text{\tiny In}}(s_{\text{\tiny In}}' ) }{D_{\text{\tiny In}}(s_{\text{\tiny In}}' )} ds_{\text{\tiny In}}'\right] , \label{EQ16}
\end{eqnarray}
where $\mathcal{N}$ is the normalization factor. From $p_{\text{s}}^{(\text{ss})}(s_{\text{\tiny In}} )$ we can evaluate the steady distribution of the nuclear spin Overhauser field $p_{\text{h}}^{(\text{ss})}(h)$. Using the same parameters~(except $\Delta_{1,0}$) in Fig.~\ref{rate01}, we plot $p_{\text{h}}^{(\text{ss})}(h)$ for different nominal detunings $\Delta_{1,0}=-0.3,~0,~0.675$ and $\pm2$~GHz in Fig.~\ref{fig2_3}(a), (b), (c) and (d), respectively. For comparison, the dashed curves denote the distribution of the nuclear spin Overhauser field in thermal equilibrium~(the mean of the Overhauser field is set to be zero, shown in Appendix \ref{AppInhomo}). One can see that for each case, the steady distribution of the nuclear spin Overhauser field is narrowed compared to that in thermal equilibrium. Also, the largest nuclear spin polarization in these four cases occur in Fig.~\ref{fig2_3}(d), where the mean of $s_{\text{\tiny In}}$ is $\sim\mp0.067$ for $\Delta_{1,0}=\pm2$~GHz, fulfilling the condition $|s_{\text{\tiny In}}|\ll1$ that is needed for deriving the Fokker-Planck equation. An $n$-times narrower $p_{\text{h}}^{(\text{ss})}(h)$ indicates $n$-times smaller an electron spin decoherence rate.\cite{Liu2007} Specifically, $\Delta_{1,0}=0.675$~GHz in Fig.~\ref{fig2_3}(c) is equal to $\Omega_{2}/2$, corresponding to the detuning of the second pump in the two-pump experiment shown in Fig.~3c of Ref.~\onlinecite{Xu2009}, where a significant enhancement of the electron spin coherence is observed. Besides a narrowed distribution of the nuclear spin Overhauser field, we also note that there are two separated peaks in $p_{\text{h}}^{(\text{ss})}(h)$ when $\Delta_{1,0}=0$, indicating bistability of the nuclear spin polarization.\cite{Braun2006,Tartakovskii2007,Kaji2008} Finally, we note that $p_{\text{h}}^{(\text{ss})}(h)$ for $|\Delta_{1,0}|$ and $-|\Delta_{1,0}|$ are symmetrical to each other about the line $h=0$, as shown in Fig.~\ref{fig2_3}(d) for illustration.

\begin{figure}
  \centering
  \includegraphics[width=3.4in]{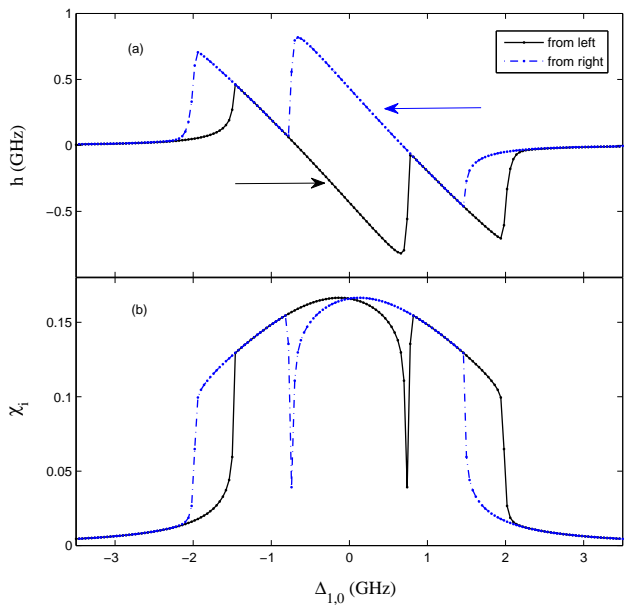}
  \caption{(Color online) (a) Nuclear spin Overhauser field, and (b) $\chi_i$ as a function of detuning $\Delta_{1,0}$ when we change the laser detuning by $0.04$~GHz at $4$~s interval. Data shown with solid~(dash-dot) curves correspond to changing $\Delta_{1,0}$ from left~(right) to right~(left). Here we use $\gamma_{\text{dep}}=0.2/s$. Other parameters except $\Delta_{1,0}$ are the same as in Fig.~\ref{fig2_3}.}
  \label{fig3}
\end{figure}

\begin{figure}
  \centering
  \includegraphics[width=3.4in]{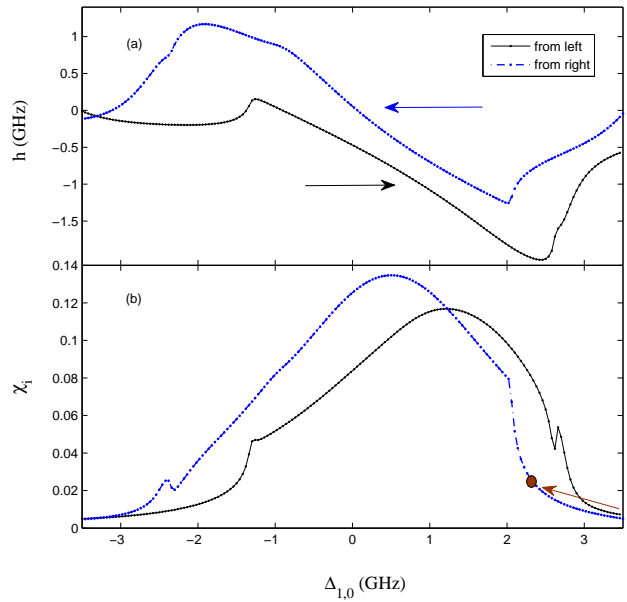}
  \caption{(Color online) Same as Fig.~\ref{fig3}. Parameters are the same as in Fig.~\ref{fig3} except $\gamma_{\text{s}}=0.22, \Delta_{2,0}=-0.8, \Omega_2=0.7$~(unit: GHz), and $\gamma_{\text{dep}}=0.02/s$.}
  \label{fig4}
\end{figure}

\begin{figure}
  \centering
  \includegraphics[width=3.4in]{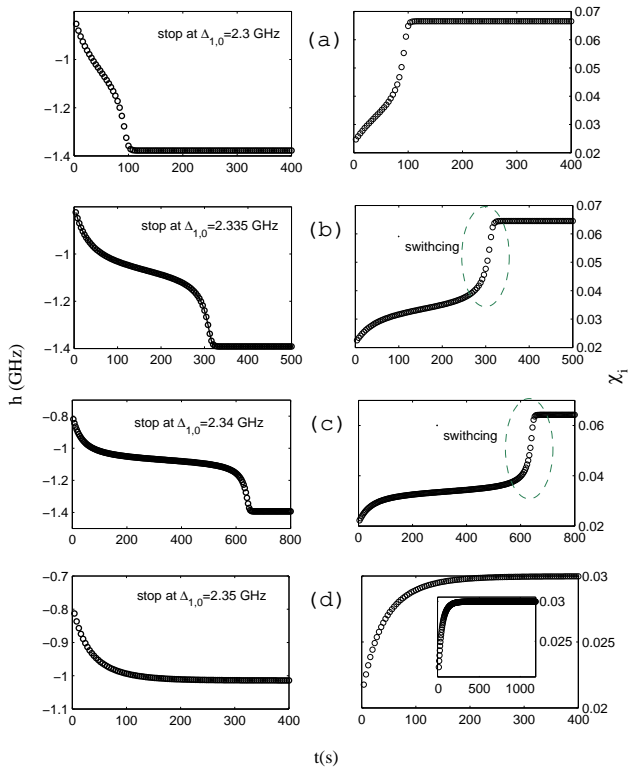}
  \caption{(Color online) Left and right panels of (a): $h$ and $\chi_i$ as a function of $t$ after stopping changing $\Delta_{1,0}$ from $3.5$~GHz step by step~[The step is $0.04$~GHz as in Fig.~\ref{fig4}] to $\Delta_{1,0}=2.3$~GHz, denoted by the arrow in Fig.~\ref{fig4}(b). (b), (c) and (d) are similar to (a) except that we stop changing $\Delta_{1,0}$ at $2.335,2.34$ and $2.35$~GHz, respectively. The inset of the right panel of (d) shows the longer time behavior of $\chi_i$. One can see that the absorption coefficient has a switching behavior in (b) and (c).}
  \label{fig5}
\end{figure}

\subsubsection{Laser spectroscopy experiments }
Here, we study systems with two different sets of parameters and show that, as they were captured, three phenomena observed experimentally in Ref.~\onlinecite{Xu2009}. The first case is for showing that the absorption curve will shift when the probe laser frequency is changed step by step. In the second case, we will show that (i) the line-shapes of $\chi_i$ for forward and backward scanning are not symmetric to each other when the pump laser is off-resonance, and (ii) there can be an abrupt switching of the probe laser absorption when we stop changing the laser frequency at a rising edge of the absorption curve. We note that these three phenomena agree with the respective experiments~(see Fig.~1, Fig.~2a, and Fig.~2c of Ref.~\onlinecite{Xu2009}, respectively).

For the first case, the parameters $\Delta_{2,0},~ \Gamma,~\gamma_{\text{s}}, ~\Omega_1,~\Omega_2$ are set as $0,~0.4,~0.06, ~0.24,~1.35$~GHz, respectively. We set $\gamma_{\text{dep}}=0.2/s$ here. We change the central frequency $\omega_1$ of the probe laser by $40$~MHz at $4$~s interval, from high to low and from low to high, corresponding to changing $\Delta_{1,0}$ from negative to positive and from positive to negative, respectively. By using Eq.~(\ref{EQ11}), the calculated Overhauser field $h$ and the absorption coefficient $\chi_i$ of the probe laser is shown in Fig.~\ref{fig3}(a) and (b), respectively. The data shown with a solid~(dash-dot) curve correspond to changing $\Delta_{1,0}$ from left to right~(from right to left), respectively. One can see that the absorption curve shifts towards the direction of changing $\Delta_{1,0}$. This result agrees with the experimental observation in Ref.~\onlinecite{Xu2009}.

For the second case, we choose a detuned pump laser with $\Delta_{2,0}=-0.8$~GHz~(this does not mean the detuning in Fig.~2a of Ref.~\onlinecite{Xu2009} is negative. We use detuning as the $x$-axis, while Fig.~2a of Ref.~\onlinecite{Xu2009} uses laser frequency), and choose the Rabi frequency of the pump laser $\Omega_2$ as $0.7$~GHz. Since the pump laser is weaker than the one used in Fig.~\ref{fig3}, we set $\gamma_{\text{s}}=0.22$~GHz, as estimated from Ref.~\onlinecite{Xu2009}~(see Fig.~1 of its supplementary information). Also, we set $\gamma_{\text{dep}}=0.02/s$, which is much smaller than the depolarization rate used in Fig.~\ref{fig3}. Interestingly, we note that in the supplementary material of Ref.~\onlinecite{Xu2009}, a smaller~(comparing with the one used for the resonant pump laser) nuclear spin depolarization rate for a detuned pump laser is also used for numerical simulation. We have chosen the parameters different from those in Fig.~\ref{fig3} in order to reproduce the experimental results in Ref.~\onlinecite{Xu2009}. Indeed, when we calculate the nuclear spin Overhauser field and the absorption coefficient of the probe laser shown in Fig.~\ref{fig4}, we can see that the line-shapes of $\chi_i$ for forward and backward scanning are not symmetric to each other. 

In the second case, we also study a switching behavior of the probe laser absorption. As indicated by the arrow in Fig.~\ref{fig4}(b), we now change $\Delta_{1,0}$ from right to left step by step until reaching $\Delta_{1,0}=2.3$~GHz, where we still keep both lasers on. At the instant we stop changing $\Delta_{1,0}$, we begin to record the laser absorption as a function of time. we present $h$ and $\chi_i$ as a function of time in the left and right panel of Fig.~\ref{fig5}(a), respectively, where we find no abrupt switching of $\chi_i$. We then repeat the same calculation as in Fig.~\ref{fig5}(a) when the point of $\Delta_{1,0}$ at which we stop changing $\Delta_{1,0}$ is set as $2.335,2.34$ or $2.35$~GHz. The respective results are shown in Fig.~\ref{fig5}(b), (c) and (d). We can see that there is an obvious switching behavior of $\chi_i$ at around $t=310$~s~(or $t=610$~s) after we stop scanning the probe laser in Fig.~\ref{fig5}(b)~[or (c)]. The switching behavior here qualitatively agrees with the experimental observation in Fig.~2c Ref.~\onlinecite{Xu2009}. Interestingly, our simulation reveals that there would be no such behavior when we stop shifting the probe laser frequency too far from the rising edge, say, at $\Delta_{1,0}=2.35$~GHz, as shown in Fig.~\ref{fig5}(d).

The hysteresis in Figs.~\ref{fig3} and \ref{fig4} and the switching effects in Fig.~\ref{fig5} result from the nonlinear feedback between the electron-hole system and the nuclear spin bath, where the feedback is controlled by the hole spin-nuclear spin noncollinear hyperfine interaction. In Appendix A, we show that the spin flip-flops due to the hyperfine interaction may be neglected for strong magnetic field. However, the feedback is much less effective in polarizing the nuclear spins if the external magnetic field is lower than, for instance, $0.1$~T. This is because the function $s_{\text{\tiny In(As)},0}$ in Eq.~(\ref{EQB09}) that is responsible for the nonlinear feedback will vanish as $\sim B$,\cite{Yang2012} thus causing the hysteresis and switching effects through the noncollinear interaction in our model to vanish. In such a case, the dynamics of the nuclear spins will be controlled primarily by the hyperfine flip-flops. It has also been noted that similar hysteresis of nuclear spin polarization in an optically pumped QD could be explained by exploiting the spin flip-flops through the electron spin-nuclear spin hyperfine interaction~[see, e.g., Refs.~\onlinecite{Braun2006,Latta2009,Tartakovskii2007,Kaji2008}].

\begin{figure}
  \centering
  \includegraphics[width=3.0in]{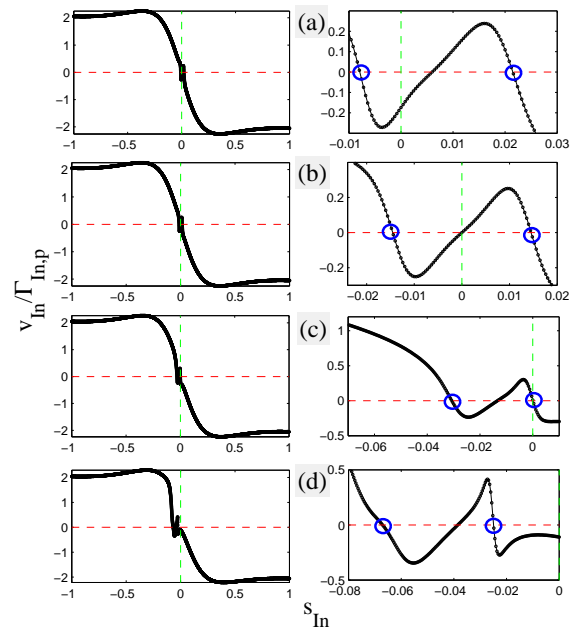}
  \caption{(Color online) $v_{\text{\tiny In}}/ \Gamma_{\text{In,p}}$ as a function of $s_{\text{\tiny In}} $. $\Delta_{1,0}=-0.3,0,0.675$ and $2$~GHz in (a), (b), (c) and (d), respectively. Other parameters are the same as in Fig.~\ref{fig2_3}. In each subfigure, the right panel zooms in the left panel, showing the crossings~[which gives the position of a local maximum of $p_{\text{s}}^{(\text{ss})}(s_{\text{\tiny In}} )$] between the curve $v_{\text{In}}/ \Gamma_{\text{In,p}}$ and the horizontal dashed line. One can see that there are two such crossings in each panel, denoted by the circles.}
  \label{fig7}
\end{figure}

\begin{figure}
  \centering
  \includegraphics[width=3.0in]{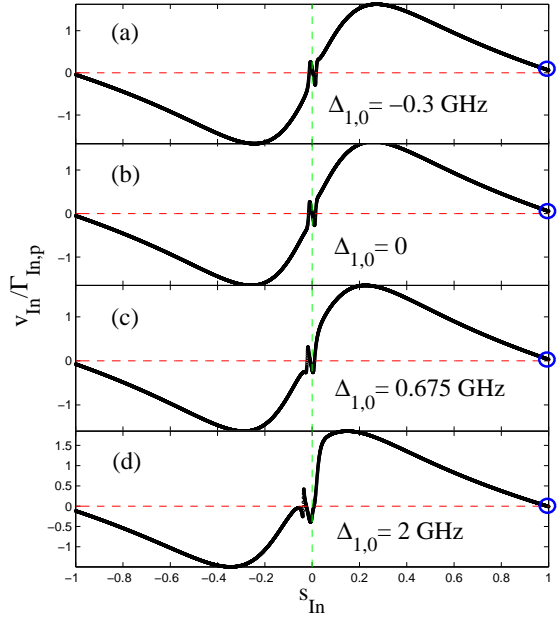}
  \caption{(Color online) Same as Fig.~\ref{fig7}, except that here $\Omega_{1}=1.35$~GHz and $\Omega_2=0.24$~GHz.}
  \label{fig8}
\end{figure}
\begin{figure}
  \centering
  \includegraphics[width=3.0in]{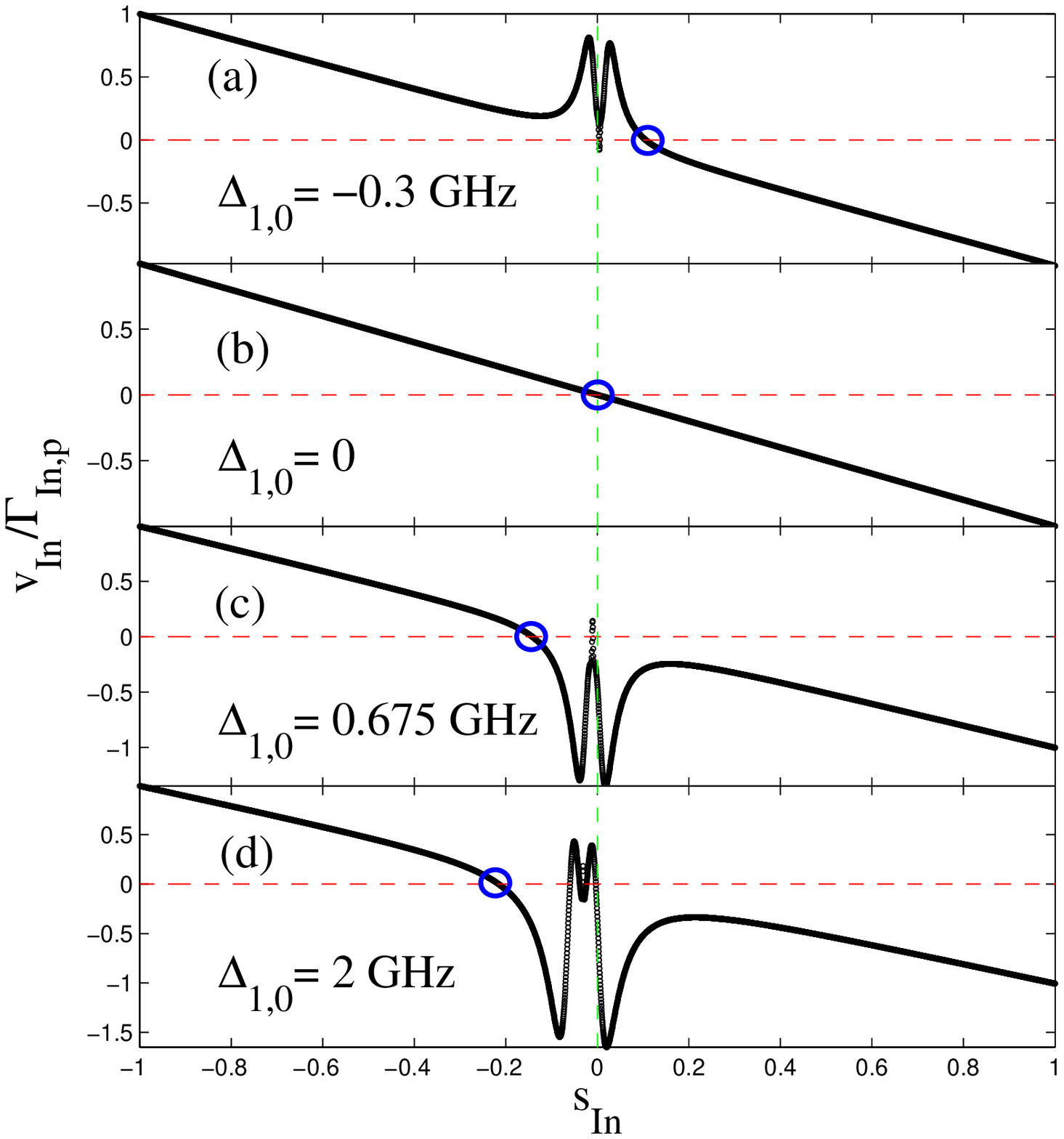}
  \caption{(Color online) Same as Fig.~\ref{fig7}, except that here $\Omega_1=\Omega_2=0.24$~GHz.}
  \label{fig9}
\end{figure}
\begin{figure}
  \centering
  \includegraphics[width=3.0in]{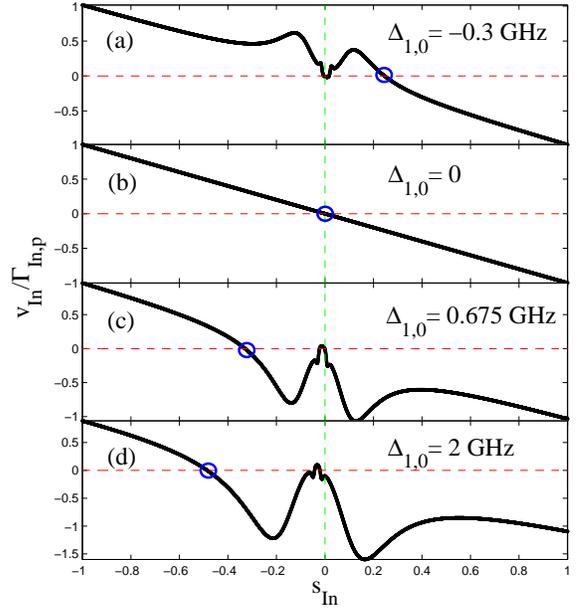}
  \caption{(Color online) Same as Fig.~\ref{fig7}, except that here $\Omega_1=\Omega_2=1.35$~GHz.}
  \label{fig10}
\end{figure}

\section{Large nuclear spin polarization}
Now, we turn to the cases in which the nuclear spin dynamics controlled by Eq.~(\ref{e09}) can induce significant nuclear spin polarization, where we cannot apply the Fokker-Planck equation from the preceding section since its derivation requires $|s_{\text{\tiny In(As)}}|\ll1$. In order to simplify the discussion, we follow Sec.~\ref{narrowedSec} and assume a system with only indium nuclear spins. 

\subsection{When $|s_{\text{\tiny In}}|\ll1$ is violated?}\label{SecVA}
Before introducing a condition where $|s_{\text{\tiny In}}|\ll1$ is not fulfilled, we first study cases where $|s_{\text{\tiny In}}|\ll1$ is satisfied and Eq.~(\ref{e014}) determines the evolution of $p_{\text{s}}(s_{\text{\tiny In}} )$. For this case, it is pointed out\cite{WenPRB} that a peak of $p_{\text{s}}^{(\text{ss})}(s_{\text{\tiny In}} )$ can usually be approximated by a Gaussian. The position $s_{\text{\tiny In}}=s_{\text{\tiny In},m}$ of such a Gaussian, i.e., a local maximum of $p_{\text{s}}^{(\text{ss})}(s_{\text{\tiny In}} )$ in Eq.~(\ref{EQ16}) is given by
\begin{eqnarray}
 \frac{v_{\text{\tiny In}}(s_{\text{\tiny In},m} ) }{D_{ \text{\tiny In}}(s_{\text{\tiny In},m} )}=0,~ \frac{\partial }{\partial s_{\text{\tiny In}}'} \frac{v_{\text{\tiny In}}(s_{\text{\tiny In}}' ) }{D_{\text{\tiny In}}(s_{\text{\tiny In}}' )}\bigg|_{s_{\text{\tiny In},m}}<0.\label{EQ17}
\end{eqnarray}
From Eq.~(\ref{EQ15}) we have $D_{  \text{\tiny In}}(s_{\text{\tiny In}} )\approx \frac{( I+1)\Gamma_{\text{In,p}}}{3NI}$, hence we can study $v_{ \text{\tiny In}} / \Gamma_{\text{In,p}} $ to find the solution of Eq.~(\ref{EQ17}). We plot $v_{ \text{In}}/ \Gamma_{\text{In,p}} $ for four different nominal detunings $\Delta_{1,0}=-0.3,0,0.675$ and $2$~GHz in Fig.~\ref{fig7}(a)-(d), i.e., the same as those in Fig.~\ref{fig2_3}(a)-(d). From Fig.~\ref{fig7}, one finds that there are two solutions~(denoted by the small circles) to Eq.~(\ref{EQ17}) for each case, hence there should be two peaks of $p_{\text{s}}^{(\text{ss})}(s_{\text{\tiny In}} )$. Nevertheless, we note that for the two local maxima of $p_{\text{s}}^{(\text{ss})}(s_{\text{\tiny In}} )$ for $\Delta_{1,0}=0.675$ or $2$~GHz, one peak is much lower than the other peak in Fig.~\ref{fig2_3}(c) or (d) that it is not visible in Fig.~\ref{fig2_3}~[mathematically, taking Fig.~\ref{fig2_3}(d) as an example, for the two deformed `triangles' formed by the solid curve and the horizontal dashed line, the area of the triangle pointing down is larger than that of the one which points up]. All local maxima of $p_{\text{s}}^{(\text{ss})}(s_{\text{\tiny In}} )$ given by Fig.~\ref{fig2_3} occur with $|s_{\text{\tiny In}}|\ll1$, indicating that only small nuclear spin polarization can be generated. This is consistent with the condition for deriving the Fokker-Planck equation.

Now we consider the case where $|s_{\text{\tiny In}}|\ll1$ is not fulfilled. As an example, we exchange the intensities of the pump and probe lasers. If we {\it assume} $|s_{\text{\tiny In}}|\ll1$, we can still derive the same Fokker-Planck equation Eq.~(\ref{EQ09}). In this formalism, we can also assume that the position for a local maximum of $p_{\text{s}}^{(\text{ss})}(s_{\text{\tiny In}} )$ is given by a solution to Eq.~(\ref{EQ17}). Again, we plot $v_{ \text{\tiny In}}(s_{\text{\tiny In}} )/ \Gamma_{\text{In, p}} $ in Fig.~\ref{fig8}. One finds that there is always a solution to Eq.~(\ref{EQ17}) at $s_{\text{\tiny In}}\sim1$. If one naively tries to calculate $p_{\text{s}}^{(\text{ss})}(s_{\text{\tiny In}} )$ in any of the four cases in Fig.~\ref{fig8}, the peak $\sim 1$ dominates and the mean of $s_{\text{\tiny In}}$ is almost $1$. On the other hand, if we use Eq.~(\ref{EQ11}) to study the evolution of the mean of nuclear spin polarization, strong polarization results. This contradicts the condition $|s_{\text{\tiny In}}|\ll1$.

When we set $\Omega_1=\Omega_2$ and choose weak pumping, e.g., $\Omega_1=\Omega_2=0.24$~GHz,  the solution to Eq.~(\ref{EQ17}) at $s_{\text{\tiny In}}\sim-0.2$, as seen from Fig.~\ref{fig9}(d), is already much farther from $0$ than the corresponding solution shown in Fig.~\ref{fig7}(d). When both lasers are strong, e.g., $\Omega_1=\Omega_2=1.35$~GHz, Fig.~\ref{fig10} shows that the solutions to Eq.~(\ref{EQ17}) can be even further from zero in Fig.~\ref{fig10}(c) and (d), thus indicating the failure of the Fokker-Planck equation.

In conclusion, the dynamic nuclear spin polarization in a system illustrated in Fig.~\ref{system01} will obey (violate) $|s_{\text{\tiny In(As)}}|\ll1$ when $\Omega_1\ll\Omega_2$~($\Omega_1\gg\Omega_2$) and $\Delta_{2,0}=0$. In any other case the condition $|s_{\text{\tiny In(As)}}|\ll1$ may fail to hold, depending on the Rabi frequencies and the detunings of the two laser beams.

\subsection{Numerical study when $|s_{\text{\tiny In}}|\sim1$}
Here, the condition for the derivation of a Fokker-Planck equation breaks down, but we can still solve Eq.~(\ref{e09}) for the dynamics of the nuclear spin Overhauser field when $|s_{\text{\tiny In}}|\sim1$. We study an artificial QD with only $40$ nuclear spin $J=\frac{3}{2}$ by the Green's function Monte Carlo~(GFMC).\cite{Hetherington1984,PhysRevB.57.11446} We assume that there are $20$ gallium and $20$ arsenic nuclear spins. The gyromagnetic ratio $g_{j}\mu_{\text{\tiny N}}=0.0093$~GHz$/$T and hyperfine constant $\mathcal{A}_{j}=10.6$~GHz, are estimated by taking the average among $^{69}$Ga, $^{71}$Ga, $^{75}$As (the natural abundances for these three are $60.1\%,39.9\%$ and 1, respectively\cite{easyspin}) for the gyromagnetic ratios and hyperfine constants, where the hyperfine constants for gallium and arsenic nuclear spins are taken from Refs.~\onlinecite{Liu2007,Braun2006}.

A detailed review of the numerical method used here can be found in Ref.~\onlinecite{tianfuthesis}. In order to implement GFMC simulation, we convert Eq.~(\ref{e09}) into the form of a first-order differential equation,
\begin{eqnarray}
  \frac{d}{dt}p_n(t) &=& Ap_n(t),
\end{eqnarray}
where $p_n(t)$ is a vector, each of whose elements representing the population of nuclear spins on one of the $d=(2J+1)^{40}$ different states. The $d\times d$ matrix $A$ determines the dynamics of the nuclear spin population. From Eq.~(\ref{e09}), one can prove that each off-diagonal element of $A$ is non-negative. Nonetheless, the diagonal element $A_{kk}$ is given by
\begin{eqnarray}
A_{kk}&=& -\left\langle k\left| \sum_j(J^2+J-\hat{J}_j^{x2} -\hat{J}_j^x ) \hat{W}_{j,+} \right.\right.\nonumber\\
&&\left.\left.+ \sum_j(J^2+J-\hat{J}_j^{x2} +\hat{J}_j^x ) \hat{W}_{j,-} \right|k\right\rangle,
\end{eqnarray}
which is non-positive. Here, $ \hat{W}_{j,\pm}$ is $\hat{W}_{\alpha_j,\pm}$ defined in Eq.~(\ref{EQ04}), and $|k\rangle$ represents one of the $d$ nuclear spin states. Most of the eigenvalues of the matrix $A$ are negative~(except one being equal to zero, which we tested by exactly diagonizing $A$ for smaller systems). The eigenvector of $A$ corresponding to the biggest eigenvalue of $A$ gives the steady population of the nuclear spins. In GFMC, we can study properties of the eigenvector corresponding to the biggest eigenvalue of the matrix $A$. To do this, we first choose a positive constant and add this constant to every diagonal element of $A$, shifting its spectrum up by making each of its diagonal matrix elements positive. This does not alter the result of the GFMC simulation.\cite{PhysRevB.57.11446} At the start of the simulation, we randomly set the state of each nuclear spin, which is equivalent to assuming that the temperature for the nuclear spins is infinitely high. During the simulation, we run $M$ independent sets of simulations simultaneously for $i_M$ Monte Carlo sampling. After each step of sampling, we record the distribution $p_{\text{s}}(s)$ of the nuclear spin polarization $s\equiv \frac{1}{NJ}\sum_j \langle \hat{J}_j^x\rangle$ calculated from the $M$ different nuclear spin states. After a certain step $i_M'$ the nuclear spin ensemble of these $M$ states reaches equilibrium, and the distribution of $s$ is calculated by analyzing the data from the $i_M'$ step to the $i_M$ step.
\begin{figure}
  \centering
  \includegraphics[width=2.9in]{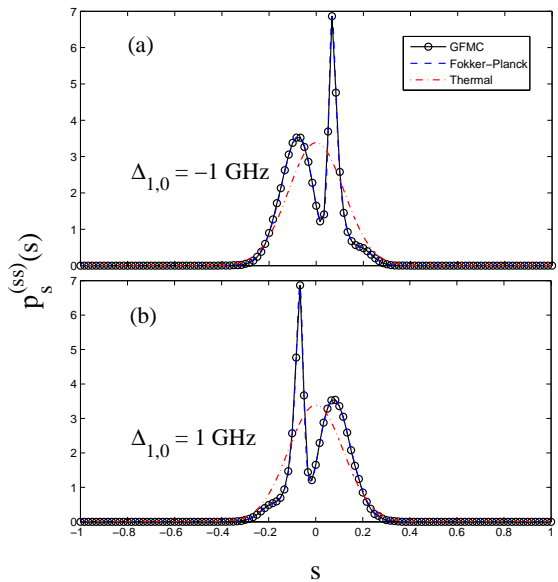}
  \caption{(Color online) Steady distribution of the nuclear spin polarization $s$ for a system with $40$ spin-$\frac{3}{2}$, calculated from the Green's function Monte Carlo simulation~(solid curve with circles) and Fokker-Planck equation~(dashed curve), respectively. Here the nuclear spin gyromagnetic ratio is $g_{j}\mu_{\text{\tiny N}}=0.0093$~GHz$/$T, and the hyperfine constant is $\mathcal{A}_{j}=10.6$~GHz. $\Delta_{1,0}=-1$ and $1$~GHz in (a) and (b), respectively. Other parameters are the same as in Fig.~\ref{rate01}. For comparison, we have shown the result for a thermal nuclear spin state by dash-dot curve.}
  \label{fig11}
\end{figure}

\begin{figure}
  \centering
  \includegraphics[width=2.9in]{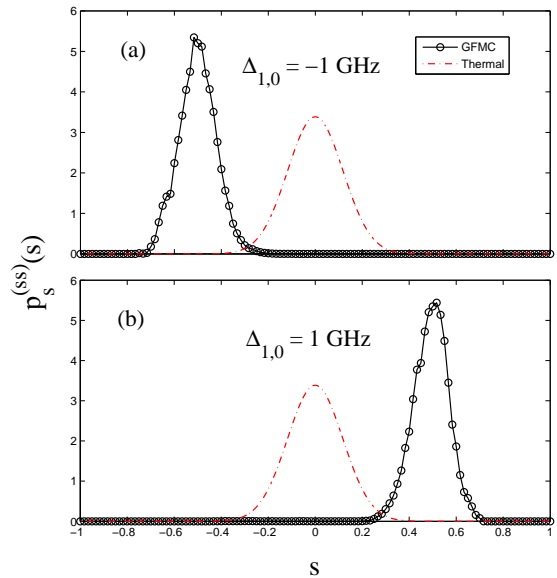}
  \caption{(Color online) The solid curves with circles and the dash-dot curves mean the same as the counterparts in Fig.~\ref{fig11} except that they are calculated with $\Omega_{1}=1.35$~GHz and $\Omega_2=0.24$~GHz. }
  \label{fig12}
\end{figure}

In order to test the GFMC in our model, we can first consider the case $\Omega_{1(2)}=0.24(1.35)$~GHz, in which case the nuclear spins polarize weakly when $N=9500$, as shown in Fig.~\ref{fig7}. We assume that the condition $|s|\ll1$ holds when we consider a small system with only $N=40$ nuclear spins. In this case, we can use both the Fokker-Planck equation and the Monte Carlo simulation to calculate the steady distribution of the nuclear spin polarization $s$. The reason that we may still use the Fokker-Planck equation is that the condition $\frac{1}{NJ}\ll1$ for deriving the Fokker-Planck equation~(see Appendix~\ref{AppendixFP}) still holds when $N=40$. In Fig.~\ref{fig11}, we present the steady distribution $p_{\text{s}}^{(ss)}(s)$ of $s$ for a system containing $40$ spin-$\frac{3}{2}$ for $\Delta_{1,0}=-1$ and $1$~GHz in (a) and (b), respectively. We calculate $p_{\text{s}}^{(ss)}(s)$ from both GFMC and Fokker-Planck equation. For the former, we set $M=200, i_M=10^5$ and $i_M'=10^4$ in the GFMC study. We also use Eq.~(\ref{EQ16}) to calculate $p_{\text{s}}^{(ss)}(s)$ for a comparison. From Fig.~\ref{fig11}, one can see that the results from GFMC and Fokker-Planck equation agree very well with each other. Another feature in Fig.~\ref{fig11} is that the distributions of $s$ for $\Delta_{1,0}=-1$ and $1$~GHz are symmetric to each about the line $s=0$.

We then consider the case $\Omega_{1(2)}=1.35~(0.24)$~GHz. From Fig.~\ref{fig8}~[the curve $v_{ \text{\tiny In}}/ \Gamma_{\text{In,p}}$ for the present case is different from that in Fig.~\ref{fig8}, yet also indicates that $s\ll1$ does not hold], we know that the criterion for using the Fokker-Planck equation fails to hold. But the Monte Carlo simulation is still valid. We sample $M=200$ simulations simultaneously. However, we found that the state represented by these samples does not reach equilibrium~[i.e., $p_{\text{s}}(s)$ is going up or down for a given $s$ if we continue the sampling] when we run only $i_M'=10^4$ simulations, as above. So we set $i_M=10^6$ here. Now $s$ is analyzed by the data generated from the $i_M'=91\times 10^4$th to the $i_M$th step. In this case, we found that the Monte Carlo simulation reaches equilibrium~(we do not mean that at least $91\times10^4$ sampling is needed to reach equilibrium). The distribution of $s$ from this Monte Carlo simulation is shown in Fig.~\ref{fig12}. One can see that the distribution of the nuclear spin polarization is peaked at $s\approx\pm0.5$ when $\Delta_{1,0}=\pm1$~GHz, and it is narrower compared to the thermal nuclear spin state without optical pumping. We note that the distributions of $s$ for $\Delta_{1,0}=-1$ and $1$~GHz are almost symmetric~(except each of them has a small kink on the left shoulder) to each other about the line $s=0$, similar to that in Fig.~\ref{fig11}. Because the peak of the nuclear spin polarization distribution is shifted from $0$ to  $\sim\pm0.5$, we conclude that strong nuclear spin polarization can be generated by optical pumping on single QDs.

\section{Conclusions}

We have studied dynamic nuclear spin polarization in a strained self-assembled QD pumped by two narrow-linewith continuous wave lasers, where the noncollinear hole spin-nuclear spin hyperfine interaction is assumed to be responsible for nuclear spin polarization. We found that the nuclear spins can be polarized to a degree that is either small or large, depending on the intensities and central frequencies of the lasers. 

We first study the experiments in Ref.~\onlinecite{Xu2009}, where only weak nuclear spin polarizations were observed. In this case, we derive a Fokker-Planck equation for the time evolution of the probability of nuclear spin polarization, based on which we show that the distribution of the nuclear spin Overhauser field can be narrowed, as well as other phenomena, including that the absorption curve of the probe laser shifts when we change the probe laser frequency. When large nuclear spin polarization can be generated by optical pumping, the condition to derive a Fokker-Planck equation breaks down. Then we use the GFMC simulation to directly solve the steady distribution of the nuclear spin Overhauser field. Indeed, we find that a large nuclear spin polarization up to $50\%$ can be generated for a small `QD' containing $40$ spin-$\frac{3}{2}$. Note that a $50\%$ nuclear spin polarization is close to the large nuclear spin polarization observed in experiments by optical pumping\cite{Gammon2001,Urbaszek2007,Skiba-Szymanska2008,Chekhovich2010} or other controls\cite{Gunnar-Petersen:2012fk} in QDs.

Throughout this paper we have assumed that the dynamic nuclear spin polarization results from the hole spin-nuclear spin noncollinear hyperfine interaction in a self-assembled QD under optical pumping. When the quadrupole interaction\cite{Abragam1983} between the electron and nuclear spins is strong compared to the Fermi-contact hyperfine interaction in the QD, there will be an effective noncollinear interaction\cite{Huang2010,Hogele2012} between the electron spin and nuclear spins which can play a similar role in polarizing the nuclear spins as studied in this paper. We note that our method is applicable when nuclear spin polarization is mainly controlled by the effective electron spin-nuclear spin noncollinear interaction.

\section*{Acknowledgments}
This work was supported by the US ARO-MURI Grant No. W911NF0910406 and NSF Grant No. PHY1104446. The author thanks L. J. Sham and Xiaodong Xu for helpful discussions, and Tiamhock Tay for introducing him to the Green's Function Monte Carlo method.

\appendix{}
\section{Steady state of the electron-hole system}
\label{steadystate}
This appendix gives the steady-state solution to the reduced density matrix of the electron-hole system in an adiabatic nuclear spin Overhauser field $h$. This steady state is determined by $\hat{H}_{\text{eh}}$ and the longitudinal part of the hyperfine interaction $\hat{H}_{\text{\tiny HI}}$ in Eq.~(\ref{e01}).

 First of all, it is helpful to use a rotating frame to show that the optically pumped electron-hole system in Fig.~\ref{system01} is an effective three-level system. The Hamiltonian for the electron-hole system includes three parts, namely the Zeeman energy $\hat{H}_{ \text{eh-b}}$ in an external magnetic field, the control $\hat{H}_{ \text{eh-l}}$ by the electromagnetic fields and the channel of photon exchange $\hat{H}_{ \text{eh-q}}$ with the electromagnetic vacuum,
  \begin{eqnarray}
    \hat{H}_{ \text{eh}}&=&  \hat{H}_{ \text{eh-b}}+ \hat{H}_{ \text{eh-l}}+ \hat{H}_{ \text{eh-q}}.
  \end{eqnarray}
\begin{widetext}
  The Zeeman term is
  \begin{eqnarray}
    \hat{H}_{ \text{eh-b}}&=&E_x |x+\rangle\langle x+|+(E_{x}+g_{\text{e}}\mu_{\text{\tiny B}}B) |x-\rangle\langle x-|
    +E_T |T+\rangle\langle T+| +(E_T+g_{\text{h}}\mu_{\text{\tiny B}}B) |T-\rangle\langle T-| ,
  \end{eqnarray}
  where $E_{x}$ and $E_T$ are the eigenenergies of the eigenstates $|x+\rangle$ and $|T+\rangle$. For the system in Fig.~\ref{system01}, the semiclassical Hamiltonian of the control by the electromagnetic fields on the electron-hole system in dipole approximation\cite{Scully97} is
  \begin{eqnarray}
    \hat{H}_{ \text{eh-l}}&=&\left( \frac{\Omega_1}{2} (e^{i\omega_1 t}+e^{-i\omega_1 t})
      |x-\rangle   +
      \frac{\Omega_2}{2}(e^{i\omega_2 t}+e^{-i\omega_2 t})
      |x+\rangle\right)\langle T-|\nonumber\\
      &&+\left( \frac{\Omega_1}{2} (e^{i\omega_1 t}+e^{-i\omega_1 t})
      |x+\rangle   +
      \frac{\Omega_2}{2} (e^{i\omega_2 t}+e^{-i\omega_2 t})
      |x-\rangle\right)\langle T+|+    \text{H.c},
  \end{eqnarray}
  where $\omega_{1(2)}$ is the central frequency of the two coherent laser beams. Here, we assume that the matrix element of the electric dipole moment\cite{Liu2010} $e\langle x+|x(y)|T\pm\rangle$ is equal to  $e\langle x-|x(y)|T\mp\rangle$ ($e$ is the elementary charge), hence the Rabi frequency for the transition $|x\mp\rangle\rightarrow |T+\rangle$ is equal to that for $|x\pm\rangle\rightarrow |T-\rangle$.
  Because the two laser beams are almost resonant with the two transitions $|x\pm\rangle\rightarrow |T-\rangle$, we use a rotating frame with 
  \begin{eqnarray}
    R &=&(E_T+g_{\text{h}}\mu_{\text{\tiny B}}B-\omega_1)|x- \rangle\langle x-|+ (E_T+g_{\text{h}}\mu_{\text{\tiny B}}B-\omega_2)|x+ \rangle\langle x+|+( E_T+g_{\text{h}}\mu_{\text{\tiny B}}B)|T-\rangle\langle T-|\nonumber\\&&+ E_T|T+\rangle\langle T+|,
  \end{eqnarray}
  to eliminate the obvious time dependence in $\hat{H}_{ \text{eh}}$ for the transition $|x\pm\rangle\rightarrow |T-\rangle$(we use the subscript $r$ to denote the rotating frame),
  \begin{eqnarray}
    \hat{H}_{\text{eh,r}} &=&  e^{iRt} \hat{H}_{\text{eh}}  e^{-iRt}-R\nonumber\\
&\approx& -\Delta_{1,0}|x-\rangle\langle x-| -\Delta_{2,0} |x+\rangle\langle x+| +\left( \frac{\Omega_1}{2} 
      |x-\rangle\langle T-|+ 
      \frac{\Omega_2}{2}
      |x+\rangle\langle T-|+    \text{H.c}\right) +e^{iRt} \hat{H}_{\text{eh-q}}  e^{-iRt},\label{EQA5}\\
    \Delta_{1,0} &=&E_T+g_{\text{h}}\mu_{\text{\tiny B}}B-( E_{x}+g_{\text{e}}\mu_{\text{\tiny B}}B+\omega_1),~~\Delta_{2,0} =E_T+g_{\text{h}}\mu_{\text{\tiny B}}B-(E_x+\omega_2),\nonumber
  \end{eqnarray}
\end{widetext}
where we have ignored the rapidly oscillating terms with phase term exp$[\pm i(\omega_1+\omega_2)t]$ or exp$[\pm i(g_{\text{e}}\pm g_{\text{h}})\mu_{\text{\tiny B}}Bt]$ since $\omega_{1(2)}\sim3\times10^5$~GHz,\cite{Xu2009} and $|(g_{\text{e}}\pm g_{\text{h}})\mu_{\text{\tiny B}}B|>13$~GHz~[with $B=-2.64$~T and $g_{\text{e(h)}}=0.49~(-0.13)$], much larger than $|\Delta_{1(2)}|$ used in the main text and other parameters in the Hamiltonian. From Eq.~(\ref{EQA5}), we can see that the effective electron-hole system involves $|x\pm\rangle$ and $|T-\rangle$, hence can be called a $\Lambda$ system~($\Lambda$S).

When we apply the rotating frame transformation to the transverse part of $\hat{H}_{\text{\tiny HI}}$, we have $a_{\text{e},j} \hat{S}_{\text{e}}^\pm \hat{I}_j^\mp \rightarrow a_{\text{e},j}\hat{S}_{\text{e}}^\pm \hat{I}_j^\mp e^{\pm it(\Delta_{2,0} - \Delta_{1,0}-g_{\text{e}}\mu_{\text{\tiny B}}B )}$. Here $g_{\text{e}}\mu_{\text{\tiny B}}B\approx18$~GHz, hence $ e^{\pm it(\Delta_{2,0} - \Delta_{1,0}-g_{\text{e}}\mu_{\text{\tiny B}}B )}$ is rapidly oscillating when $\Delta_{1,0}$ and $\Delta_{2,0}$ are set as in the main text. So the transverse part of the electron spin-nuclear spin hyperfine interaction can be neglected. We note that a Schrieffer-Wolff transformation\cite{PhysRev.90.297,PhysRev.149.491} on the transverse part of $\hat{H}_{\text{\tiny HI}}$ may give rise to a nonlinear term $\hat{S}_{\text{e}}^x  \sum_{j,j'\neq j}\frac{a_{\text{e},j}a_{\text{e},j'}}{ 2g_{\text{e}}\mu_{\text{\tiny B}}  B}\hat{I}_j^+ \hat{I}_{j'}^-$, which also contributes to nuclear spin dynamics.\cite{Latta2011} However, one can evaluate~\cite{xfshi13} and show that the pumping of nuclear spins through this nonlinear term is much slower than that through the noncollinear term $\sim \hat{S}_{\text{h}}^x\hat{I}_j^y $ for the specific system studied in this paper. For such case, we shall assume that the transverse part of electron spin-nuclear spin hyperfine interaction can be neglected. Finally, since $a_{\text{h},j}/a_{\text{e},j}\sim-0.1$~(see Refs.~\onlinecite{Chekhovich2011,Fallahi2010}) and the strength of the coupling between $|T+\rangle$ and $|T-\rangle$ given by a mean field treatment of $\frac{2|\beta|a_{\text{h},j}}{\sqrt{3}}\hat{S}_{\text{h}}^y\hat{I}_j^x\sin\delta$ is weak compared to the Rabi frequencies and decay rate of the trion in this paper, we further neglect the transverse part of the hole spin-nuclear spin hyperfine interaction. Equation~(\ref{e02}) then becomes
\begin{eqnarray}
  \hat{H}_\text{\tiny HI}&\approx&
  \sum_j\left[a_{\text{e},j}\hat{S}_{\text{e}}^x\hat{I}_j^x+ \frac{2|\beta|a_{\text{h},j}  }{\sqrt{3}} \hat{S}_{\text{h}}^x(\hat{I}_j^x\cos\delta+\hat{I}_j^y\sin\delta)\right],\nonumber\\
\label{eA06}
\end{eqnarray}
where we have used $1+ |\beta|^2\approx1$ when $|\beta|=0.2$.

We perform the mean field approximation on the longitudinal part of the electron~(hole) spin-nuclear spin hyperfine interaction. This gives an effective nuclear spin field, i.e., the Overhauser field that shifts the Zeeman energy of the electron~(hole) spin,
\begin{eqnarray}
  h&=&\sum_j a_{\text{e},j}\langle\hat{I}_j^x\rangle,  \label{eq07}\\
  h_{\text{h}}&\approx&\sum_j \frac{2|\beta| a_{\text{h},j}}{\sqrt{3}}\langle\hat{I}_j^x\rangle \cos\delta .  
 \end{eqnarray}
Since $|\beta|\sim0.2$~(taken from Ref.~\onlinecite{Xu2009}) and $a_{\text{h},j}/a_{\text{e},j}\sim-0.1$, we neglect $h_{\text{h}}$ but keep the nuclear spin Overhauser field $h$ via the electron spin-nuclear spin hyperfine interaction. As a result, the actual detuning between the photon energy of the laser and the energy cost in relevant transitions is  
  \begin{eqnarray}
    \Delta_{1} &=&    \Delta_{1,0}+\frac{ h}{2},~~    \Delta_{2} =   \Delta_{2,0}-\frac{ h}{2}.
  \end{eqnarray}

Next, we solve the steady-state reduced density matrix of the $\Lambda$S in an adiabatic nuclear spin Overhauser field $h$. The last term in Eq.~(\ref{EQA5}) induces spontaneous decay of the trion level, which we incorporate in the optical Bloch equation in the Lindblad form,\cite{Fleischhauer2005}
\begin{eqnarray}
  \frac{d\hat{\rho}_\Lambda}{dt}   &=& -i[\hat{H}_{\Lambda\text{S},\text{r}},\hat{\rho}_{\Lambda}]+\Gamma\sum_{k=1}^2 \left[\hat{G}_k\hat{\rho}_{\Lambda} \hat{G}_k^\dag - \frac{1}{2}\left\{\hat{G}_k^\dag \hat{G}_k, \hat{\rho}_{\Lambda} \right\}\right]\nonumber\\
  && + \gamma_s\sum_{k=3}^4 \left[\hat{G}_k\hat{\rho}_{\Lambda} \hat{G}_{k}^\dag - \frac{1}{2}\left\{\hat{G}_{k}^\dag \hat{G}_k,\hat{\rho}_{\Lambda} \right\}\right],\label{EQA10}
\end{eqnarray}
where $\{a,b\}=ab+ba$, and
\begin{eqnarray}
\hat{H}_{\Lambda\text{S},\text{r}}&=&  -\Delta_{1}|x-\rangle\langle x-|-\Delta_{2} |x+\rangle\langle x+|\nonumber\\
&&+\left( \frac{\Omega_1}{2} 
      |x-\rangle\langle T-|+ 
      \frac{\Omega_2}{2}
      |x+\rangle\langle T-|+    \text{H.c}\right),\nonumber\\
  \hat{G}_{1,2}&=&|x\pm \rangle\langle T-|,~\hat{G}_{3,4}=|x\pm \rangle\langle x\pm|,
\end{eqnarray}
and the $\Lambda$-system reduced density matrix $\hat{\rho}_{\Lambda}$ in the basis constructed from the vectors $|T-\rangle,|x+\rangle$ and $|x-\rangle$ is
\begin{eqnarray}
 \hat{\rho}_{\Lambda}&=& \left( \begin{array}{ccc}\rho_{\overline{T},\overline{T}}& \rho_{\overline{T},x} & \rho_{\overline{T},\overline{x}}\\
    \rho_{x, \overline{T}}& \rho_{x,x}& \rho_{x,\overline{x} }\\
    \rho_{\overline{x},\overline{T}}& \rho_{\overline{x},x}& \rho_{\overline{x},\overline{x} }\end{array} \right).
\end{eqnarray}
The condition Tr$\hat{\rho}_{\Lambda}=1$ allows us to eliminate one matrix element, say, $\rho_{\overline{x},\overline{x}}|x-\rangle\langle x-|$ in  $\hat{\rho}_{\Lambda}$ and rearrange its other eight matrix elements into one column,
\begin{eqnarray}
  \overline{\rho}_{\Lambda}&=& (\rho_{\overline{T},\overline{T}},\rho_{x,x}, \rho_{x, \overline{T}} ,\rho_{\overline{x}, \overline{T}}, \rho_{x,\overline{x} }, \rho_{\overline{T},x}, \rho_{\overline{T},\overline{x}}, \rho_{\overline{x} ,x})^T,\nonumber
\end{eqnarray}
where $T$ denotes transposing a matrix. Then Eq.~(\ref{EQA10}) becomes,
\begin{eqnarray}
  i \frac{d}{dt} \overline{ \rho}_{\Lambda}(t) &=& \mathcal{M}  \overline{\rho}_{\Lambda}(t)+X.
\end{eqnarray}
Here,
 \begin{eqnarray}
  X &=&(0,~0,~0,~-\frac{\Omega_1}{2} ,~0,~0,~\frac{\Omega_1}{2},~0   ),\label{eA7}
\end{eqnarray}
and
\begin{widetext}
  \begin{eqnarray}
    \mathcal{M}  &=&\left(\begin{array}{cccccccc}
      -i\Gamma_1 & 0 & \frac{\Omega_2}{2} & \frac{\Omega_1}{2} & 0 & -\frac{\Omega_2}{2} & -\frac{\Omega_1}{2} &0\\
      \frac{i}{2}\Gamma_1 & 0 & -\frac{\Omega_2}{2} & 0& 0 &\frac{\Omega_2}{2}& 0 & 0\\
      \frac{\Omega_2}{2} &- \frac{\Omega_2}{2} & -\Delta_2-i\Gamma_2 & 0 &-\frac{\Omega_1}{2}& 0 & 0 &0\\
      \Omega_1 & \frac{\Omega_1}{2} & 0 & -\Delta_1-i\Gamma_2 & 0 & 0 & 0&-\frac{\Omega_2}{2}\\
      0 & 0 & - \frac{\Omega_1}{2} & 0 &\Delta_1-\Delta_2-i \gamma_s & 0 & \frac{\Omega_2}{2} & 0\\
      -\frac{\Omega_2}{2} & \frac{\Omega_2}{2} &0 & 0 &0& \Delta_2-i\Gamma_2 & 0 &\frac{\Omega_1}{2}\\
      -\Omega_1 & -\frac{\Omega_1}{2} & 0 &0 &\frac{\Omega_2}{2} & 0 & \Delta_1-i\Gamma_2&0\\
      0 & 0& 0 & -\frac{\Omega_2}{2} &0 &  \frac{\Omega_1}{2} &0 &  \Delta_2-\Delta_1- i \gamma_{s} \end{array}
    \right),\label{eA6}
  \end{eqnarray}
\end{widetext}
where $\Gamma_1\equiv 2\Gamma$ is the relaxation rate of the trion state, $\Gamma_2=\Gamma+\frac{\gamma_s}{2}$ is the decay rate of the coherence between the trion and one of the electron spin eigenstates, $\Gamma$ is the spontaneous decay rate from the trion to each of the two electron spin states, and $\gamma_s$ is the energy-conserving dephasing rate of the electron spin in the presence of fluctuating nuclear spins. Note that we have ignored the electron spin relaxation process which is much slower than the electron spin decoherence.\cite{Khaetskii2000}

In the steady state of the $\Lambda$S, we obtain
\begin{eqnarray}
  \mathcal{M}  \overline{\rho}_{\Lambda}(t\rightarrow+\infty)+X&=&0,
\end{eqnarray}
whose solution, together with the condition Tr$\hat{\rho}_{\Lambda}=1$, gives us the reduced density matrix $\hat{\rho}_{\Lambda}^{(ss)}$ for the steady state of the $\Lambda$S.

\section{Fokker-Planck equation}
\label{AppendixFP}
In this appendix, we derive the Fokker-Planck equation starting from Eq.~(\ref{e09}). For the derivation of an analogous equation involving only one species of nuclear spins, see Ref.~\onlinecite{Yang2012}. We first multiply the  Kronecker delta function $\delta_{\hat{s}_{\text{In}},s_{\text{\tiny In}} }\delta_{\hat{s}_{\text{As}},s_{\text{\tiny As}}}$ on both sides of Eq.~(\ref{e09}), and then trace over the nuclear spin degrees of freedom in it, giving
\begin{widetext}
\begin{eqnarray}
  \frac{d}{dt}\text{Tr}[\hat{P}(t) \delta_{\hat{s}_{\text{In}},s_{\text{\tiny In}} }\delta_{\hat{s}_{\text{As}},s_{\text{\tiny As}}}]  &\approx &  -\sum_{j} \text{Tr} \left\{  \left[   \hat{I}_j^- ,~~ \hat{I}_j^+ \hat{W}_{\alpha_j,+}  \hat{P}(t)   \right]\delta_{\hat{s}_{\text{In}},s_{\text{\tiny In}} }\delta_{\hat{s}_{\text{As}},s_{\text{\tiny As}}}+\left[   \hat{I}_j^+  ,~~ \hat{I}_j^- \hat{W}_{\alpha_j,-}   \hat{P}(t)   \right]\delta_{\hat{s}_{\text{In}},s_{\text{\tiny In}} }\delta_{\hat{s}_{\text{As}},s_{\text{\tiny As}}}\right\},\nonumber\\\label{EQB01}
\end{eqnarray}
where Tr denotes tracing over the nuclear spin degrees of freedom. Using Eq.~(\ref{e15}), the left hand side of Eq.~(\ref{EQB01}) becomes $\frac{d}{dt}p_{\text{s}}(s_{\text{\tiny In}} ,s_{\text{\tiny As}})$ [we suppress the variable $t$ in $p_{\text{s}}$ for brevity]. Taking the sum in the first  commutator on the right hand side of Eq.~(\ref{EQB01}) as an example, the trace evaluates to
\begin{eqnarray}
  -\sum_{j} \text{Tr} \left\{  \left[   \hat{I}_j^- ,~~ \hat{I}_j^+ \hat{W}_{\alpha_j,+}  \hat{P}(t)   \right]\delta_{\hat{s}_{\text{In}},s_{\text{\tiny In}} }\delta_{\hat{s}_{\text{As}},s_{\text{\tiny As}}} \right\}&=&  -\sum_{j}\sum_k \langle k|\left\{  \left[   \hat{I}_j^- ,~~ \hat{I}_j^+ \hat{W}_{\alpha_j,+}  \hat{P}(t)   \right]\delta_{\hat{s}_{\text{In}},s_{\text{\tiny In}} }\delta_{\hat{s}_{\text{As}},s_{\text{\tiny As}}} \right\}|k\rangle,\label{EQB02}
\end{eqnarray}
where $k$ labels a specific nuclear spin state and runs over the whole Hilbert space of the nuclear spin states.

First, we evaluate the sum for $\hat{I}_j^- \hat{I}_j^+ \hat{W}_{\alpha_j,+}  \hat{P}(t)$ on the right hand side of Eq.~(\ref{EQB02}):
\begin{eqnarray}
 -\sum_{j}\sum_k \langle k|\left\{  \left[   \hat{I}_j^-  \hat{I}_j^+ \hat{W}_{\alpha_j,+}  \hat{P}(t)   \right]\delta_{\hat{s}_{\text{In}},s_{\text{\tiny In}} }\delta_{\hat{s}_{\text{As}},s_{\text{\tiny As}}} \right\}|k\rangle&=&
  -\sum_{j}\sum_k \langle k| \left[I_j(I_j+1)- \hat{I}_j^{x2}- \hat{I}_j^x\right] \hat{W}_{\alpha_j,+}  \hat{P}(t)   \delta_{\hat{s}_{\text{In}},s_{\text{\tiny In}} }\nonumber\\&&\times\delta_{\hat{s}_{\text{As}},s_{\text{\tiny As}}}  |k\rangle.\label{EQB03} 
\end{eqnarray}
Below, we label $I_{\text{In}}=I=\frac{9}{2}$ and $I_{\text{As}}=J=\frac{3}{2}$ unless otherwise specified. Now assume that the following approximation is valid when both polarizations of the indium and arsenic nuclear spins are small,\cite{Yang2012}
\begin{eqnarray}
  \left\langle  \sum_{j\in \text{In(As)}} (\hat{I}_j^{x})^2\right \rangle\approx \frac{NI_{\text{\tiny In(As)}}(I_{\text{\tiny In(As)}}+1)}{3} ,\label{EQB04}
\end{eqnarray}
where $\langle\dots\rangle$ denotes the respective expectation value, then Eq.~(\ref{EQB03}) becomes,
\begin{eqnarray}
 -\sum_{j}\sum_k \langle k|\left\{  \left[   \hat{I}_j^-  \hat{I}_j^+ \hat{W}_{\alpha_j,+}  \hat{P}(t)   \right]\delta_{\hat{s}_{\text{In}},s_{\text{\tiny In}} }\delta_{\hat{s}_{\text{As}},s_{\text{\tiny As}}} \right\}|k\rangle&\approx&
  -\sum_k \sum_{\alpha=\text{In,As}}\langle k| \left[\frac{2NI_\alpha(I_\alpha+1)}{3}-NI_\alpha\hat{s}_{\alpha}\right] \hat{W}_{\alpha,+}  \hat{P}(t)  \nonumber\\
  &&\times \delta_{\hat{s}_{\text{In}},s_{\text{\tiny In}} }\delta_{\hat{s}_{\text{As}},s_{\text{\tiny As}}}  |k\rangle\nonumber\\&=&
  - \sum_{\alpha=\text{In,As}} \left[\frac{2NI_\alpha(I_\alpha+1)}{3}-NI_\alpha\hat{s}_{\alpha}\right] W_{\alpha,+}(s_{\text{\tiny In}} ,s_{\text{\tiny As}}) p_{\text{s}}(s_{\text{\tiny In}} ,s_{\text{\tiny As}}),\nonumber\\\label{EQB05}
\end{eqnarray}
where $\hat{s}_{\alpha}$, with $\alpha=$~In~(As), is defined in Eq.~(\ref{EQ06}).

Next, we evaluate the remaining sum on the right hand side of Eq.~(\ref{EQB02}):
\begin{eqnarray}
 \sum_{j}\sum_k \langle k|\left\{  \left[    \hat{I}_j^+ \hat{W}_{\alpha_j,+}  \hat{P}(t)   \hat{I}_j^- \right]\delta_{\hat{s}_{\text{In}},s_{\text{\tiny In}} }\delta_{\hat{s}_{\text{As}},s_{\text{\tiny As}}} \right\}|k\rangle&=&
  \sum_{j\in \text{In}}\sum_k \langle k| \left[I(I+1)- \hat{I}_j^{x2}- \hat{I}_j^x\right] \hat{W}_{\alpha_j,+}  \hat{P}(t)   \delta_{\hat{s}_{\text{In}},s_{\text{\tiny In}} -a}\delta_{\hat{s}_{\text{As}},s_{\text{\tiny As}}}  |k\rangle\nonumber\\  &&+\text{ (similar term with sum over As)}\nonumber\\ &\approx&NI \left[\frac{2(I+1)}{3}-s_{\text{\tiny In}}+a \right] W_{\text{In},+}(s_{\text{\tiny In}}-a  ,s_{\text{\tiny As}}) p_{\text{s}}(s_{\text{\tiny In}}-a ,s_{\text{\tiny As}})\nonumber\\ &&+NJ \left[\frac{2(J+1)}{3}-s_{\text{\tiny As}}+b \right] W_{\text{As},+}(s_{\text{\tiny In}}  ,s_{\text{\tiny As}}-b) p_{\text{s}}(s_{\text{\tiny In}} ,s_{\text{\tiny As}}-b), \nonumber\\\label{EQB06}
   \end{eqnarray}
where $a=\frac{1}{NI}$, and $b=\frac{1}{NJ}$. 

Similar to Eqs.~(\ref{EQB05}) and (\ref{EQB06}), the sum over the second commutator on the right hand side of Eq.~(\ref{EQB01}) gives,
\begin{eqnarray}
  -\sum_{j} \text{Tr} \left\{  \left[   \hat{I}_j^+ ,~~ \hat{I}_j^- \hat{W}_{\alpha_j,-}  \hat{P}(t)   \right]\delta_{\hat{s}_{\text{In}},s_{\text{\tiny In}} }\delta_{\hat{s}_{\text{As}},s_{\text{\tiny As}}} \right\}&=& - \sum_{\alpha=\text{In,As}} NI_\alpha\left[\frac{2(I_\alpha+1)}{3}+\hat{s}_{\alpha}\right] W_{\alpha,-}(s_{\text{\tiny In}} ,s_{\text{\tiny As}}) p_{\text{s}}(s_{\text{\tiny In}} ,s_{\text{\tiny As}})\nonumber\\ & &+NI \left[\frac{2(I+1)}{3}+s_{\text{\tiny In}}+a \right] W_{\text{In},-}(s_{\text{\tiny In}}+a  ,s_{\text{\tiny As}}) p_{\text{s}}(s_{\text{\tiny In}}+a ,s_{\text{\tiny As}})\nonumber\\ &&+NJ \left[\frac{2(J+1)}{3}+s_{\text{\tiny As}}+b \right] W_{\text{As},-}(s_{\text{\tiny In}}  ,s_{\text{\tiny As}}+b) p_{\text{s}}(s_{\text{\tiny In}} ,s_{\text{\tiny As}}+b). \nonumber\\\label{EQB07}
  \end{eqnarray}
To simplify Eq.~(\ref{EQB07}), we perform series expansion up to the second order in $a$ and $b$, and obtain
\begin{eqnarray}
  -\sum_{j} \text{Tr} \left\{  \left[   \hat{I}_j^+ ,~~ \hat{I}_j^- \hat{W}_{\alpha_j,-}  \hat{P}(t)   \right]\delta_{\hat{s}_{\text{In}},s_{\text{\tiny In}} }\delta_{\hat{s}_{\text{As}},s_{\text{\tiny As}}} \right\}&\approx&NI aW_{\text{In},-}(s_{\text{\tiny In}}  ,s_{\text{\tiny As}}) p_{\text{s}}(s_{\text{\tiny In}} ,s_{\text{\tiny As}})\nonumber\\ &&+NJb W_{\text{As},-}(s_{\text{\tiny In}}  ,s_{\text{\tiny As}}) p_{\text{s}}(s_{\text{\tiny In}} ,s_{\text{\tiny As}}). \nonumber\\ &&+NI \left[\frac{2(I+1)}{3}+s_{\text{\tiny In}}+a \right] \left[a\frac{\partial W_{\text{In},-} p_{\text{s}}}{\partial s_{\text{\tiny In}}}+\frac{a^2}{2}\frac{\partial^2 W_{\text{In},-} p_{\text{s}}}{\partial s_{\text{\tiny In}}^2}\right]\nonumber\\ &&+NJ \left[\frac{2(J+1)}{3}+s_{\text{\tiny As}}+b \right] \left[a\frac{\partial W_{\text{As},-} p_{\text{s}}}{\partial s_{\text{\tiny As}}}+\frac{b^2}{2}\frac{\partial^2 W_{\text{As},-} p_{\text{s}}}{\partial s_{\text{\tiny As}}^2}\right]. \nonumber\\ \label{EQB08}
  \end{eqnarray}
Defining
  \begin{eqnarray}
\Gamma_{\text{In(As),p}}(s_{\text{\tiny In}} ,s_{\text{\tiny As}})&=&  {W}_{\text{In(As)},+}(s_{\text{\tiny In}} ,s_{\text{\tiny As}})+  {W}_{\text{In(As)},-}(s_{\text{\tiny In}} ,s_{\text{\tiny As}}) ,
  \nonumber\\
s_{\text{\tiny In(As)},0}(s_{\text{\tiny In}} ,s_{\text{\tiny As}}) &=& \left[{W}_{\text{In(As)},+}(s_{\text{\tiny In}} ,s_{\text{\tiny As}})-  {W}_{\text{In(As)},-}(s_{\text{\tiny In}} ,s_{\text{\tiny As}})\right]/ \Gamma_{\text{In(As),p}}(s_{\text{\tiny In}} ,s_{\text{\tiny As}}), \label{EQB09}
\end{eqnarray}
and using Eqs.~(\ref{EQB05}), (\ref{EQB08}) and series expansion from Eq.~(\ref{EQB06}) [similar to Eq.~(\ref{EQB08})], Eq.~(\ref{EQB01}) becomes,
\begin{eqnarray}
  \frac{d}{dt}p_{\text{s}} &\approx & \sum_{\alpha=\text{In,As}}\left\{ \Gamma_{\alpha,\text{p}} p_{\text{s}}+\frac{2(I_\alpha+1)}{3}\left[\frac{1}{2NI_\alpha}\frac{\partial^2    \Gamma_{\alpha,\text{p}}p_{\text{s}}}{\partial s_{\alpha}^2} - \frac{\partial   s_{\alpha,0} \Gamma_{\alpha,\text{p}}p_{\text{s}}}{\partial s_{\alpha}} \right]+ s_{\alpha}\left[-\frac{1}{2NI_\alpha}\frac{\partial^2    s_{\alpha,0} \Gamma_{\alpha,\text{p}}p_{\text{s}}}{\partial s_{\alpha}^2} + \frac{\partial   \Gamma_{\alpha,\text{p}}p_{\text{s}}}{\partial s_{\alpha}} \right]\right\}\nonumber\\
  &&-\sum_{\alpha=\text{In,As}}\frac{1}{NI_\alpha} \frac{\partial   s_{\alpha,0} \Gamma_{\alpha,\text{p}}p_{\text{s}}}{\partial s_{\alpha}} \nonumber\\
  &=& \sum_{\alpha=\text{In,As}}\frac{\partial   }{\partial s_{\alpha} } \left[ \frac{\partial   }{\partial s_{\alpha}} D_{\alpha} p_{\text{s}}(s_{\text{\tiny In}} ,s_{\text{\tiny As}}) -v_{\alpha}p_{\text{s}}(s_{\text{\tiny In}} ,s_{\text{\tiny As}})  \right], \label{EQB10}
  \end{eqnarray}
where we have dropped terms $\sim 1/N^2$. The drift and diffusion coefficients in Eq.~(\ref{EQB10}) are given by\cite{Carmichael1999,Yang2012,WenPRB} 
\begin{eqnarray}
 v_{\text{\tiny In(As)}} &=& -\Gamma_{\text{In(As),p}}\left[s_{\text{\tiny In(As)}}  -\frac{2( I_{\text{In(As)}}+1)}{3}s_{\text{\tiny In(As)},0}\right],\nonumber\\
 D_{\text{In(As)}} &=& \frac{\Gamma_{\text{In(As),p}}}{2NI_{\text{In(As)}}}\left[\frac{2( I_{\text{In(As)}}+1)}{3}- s_{\text{\tiny In(As)}}s_{\text{\tiny In(As)},0}\right].\label{EQB11}
\end{eqnarray}
 In the definition of $v_{\text{\tiny In(As)}}$ and $D_{\text{In(As)}}$ in Eq.~(\ref{EQB11}), the factor $\frac{2( I_{\text{In(As)}}+1)}{3}$ comes from the sum of individual nuclear spin fluctuations in Eq.~(\ref{EQB04}), and $\Gamma_{\text{In(As),p}}$ is the rate of nuclear spin flip due to the noncollinear interaction in our model. If $\Gamma_{\text{In(As),p}}$ is large, the distribution of the nuclear spin polarization evolves rapidly. The nonlinear function $s_{\text{\tiny In(As)},0}$ plays a key role in our feedback loop since the solution to $s_{\text{\tiny In(As)}}  -\frac{2( I_{\text{In(As)}}+1)}{3}s_{\text{\tiny In(As)},0}=0$ gives us the stable nuclear spin polarization~(see Sec.~\ref{SecVA}). Because $|s_{\text{\tiny In(As)}}s_{\text{\tiny In(As)},0}|\ll1$, the diffusion coefficient can be approximated as
\begin{eqnarray} 
 D_{\text{In(As)}} &\approx& \frac{\Gamma_{\text{In(As),p}}}{2NI_{\text{In(As)}}} \frac{2( I_{\text{In(As)}}+1)}{3}=\sigma_{\text{In(As),th}}^2\Gamma_{\text{In(As),p}},
 \end{eqnarray} 
 where $\sigma_{\text{In(As),th}}\equiv \sqrt{\frac{ I_{\text{In(As)}}+1}{3NI_{\text{In(As)}}}}$ is the standard deviation of nuclear spin polarization distribution for $N$ indium~(arsenic) nuclear spins at infinite temperature.\cite{Liu2007,Yang2012}  The appearance of $\sigma_{\text{In(As),th}}$ in the definition of $ D_{\text{In(As)}}$ is not surprising since $D_{\text{In(As)}}$ is associated with the diffusion process.\cite{Hannes1984}

Defining the means of the nuclear spin polarization $s_{\text{\tiny In}}$ and $s_{\text{\tiny As}}$ by
\begin{eqnarray}
\overline{s}_{\text{\tiny In}}  &=& \int \int s_{\text{\tiny In}} p_{\text{s}}ds_{\text{\tiny In}}ds_{\text{\tiny As}}, \nonumber\\
\overline{s}_{\text{\tiny As}}  &=& \int \int s_{\text{\tiny As}} p_{\text{s}}ds_{\text{\tiny In}}ds_{\text{\tiny As}},
\end{eqnarray}
and further assuming that the probability for the nuclear spins to be totally polarized is negligible, i.e., $p_{\text{s}}(s_{\text{\tiny In}} =\pm1, s_{\text{\tiny As}}) =p_{\text{s}}(s_{\text{\tiny In}} , s_{\text{\tiny As}}=\pm1) = 0$, we obtain\cite{Carmichael1999}
\begin{eqnarray}
 \frac{d}{d t} \overline{s}_{\text{\tiny In}}   &=& \int \int  v_{\text{\tiny In} }  p_{\text{s}}(s_{\text{\tiny In}} ,s_{\text{\tiny As}}) ds_{\text{\tiny In}}ds_{\text{\tiny As}}, \nonumber\\ \frac{d}{d t} \overline{s}_{\text{\tiny As}}   &=& \int \int  v_{\text{\tiny As} } p_{\text{s}}(s_{\text{\tiny As}} ,s_{\text{\tiny As}}) ds_{\text{\tiny In}}ds_{\text{\tiny As}}.
\end{eqnarray}

\end{widetext}

\section{Inhomogeneous broadening}
\label{AppInhomo}
The nuclear spin Overhauser field obeys a Gaussian distribution when the nuclear spins are in thermal equilibrium,\cite{Merkulov2002, Liu2007}
\begin{eqnarray}
  p_{\text{h}}(h) &\approx&   \frac{1}{\sqrt{2\pi} \Gamma_2^\ast  }e^{ -(h-\overline{h})^2/(2\Gamma_2^{\ast2} ) }, \label{eD1}\\
  \overline{h}&=&\mathcal{A}_{\text{In}}I\overline{s_{\text{\tiny In}}}+ \mathcal{A}_{\text{As}}J\overline{s_{\text{\tiny As}}} , \nonumber\\
 \Gamma_2^{\ast2} &\approx&\frac{\mathcal{A}_{\text{In}}^2I(I+1)+ \mathcal{A}_{\text{As}}^2J(J+1)}{3N},\nonumber
\end{eqnarray}
when $|s_{\text{\tiny In(As)}}|\ll1$ for temperature $\sim 5$~K in typical experiments.\cite{Latta2009,Xu2009} Here $\overline{h}$ and $\overline{s_{\alpha}}$ are the averages of $h$ and $s_{\alpha}$, respectively. The electron spin decoherence $L(t)$ is dominated by the inhomogeneous broadening of the nuclear spin Overhauser field,\cite{Yao2006}
\begin{eqnarray}
  L(t) &=& \int p_{\text{h}}(h)e^{-i(g_{\text{e}}\mu_{\text{\tiny B}}B+h)t   }dh.
\end{eqnarray}
 For a thermal nuclear spin bath of  Eq.~(\ref{eD1}), we have a pure Gaussian decay,\cite{Merkulov2002,Yao2006,Liu2007}
\begin{eqnarray}
  L(t) &=& e^{-i(g_{\text{e}}\mu_{\text{\tiny B}}B+\overline{h})t-\left(t/T_2^\ast\right  )^2     }, 
\end{eqnarray}
where $T_2^{\ast}=\sqrt{2}/\Gamma_2^\ast$. Both indium and arsenic nuclear spins contribute to $\Gamma_2^\ast$ in a QD with $N$ indium and $N$ arsenic nuclear spins. Under the condition of $|s_{\text{\tiny In(As)}}|\ll1$ and further ignoring the arsenic nuclear spins in the QD, it is useful to rescale the hyperfine constant of indium nuclear spins so that the new system with only $N$ indium nuclear spins gives a $\Gamma_2^{\ast}$ that is the same as in the original InAs system. This is achieved by writing
\begin{eqnarray}
  \mathcal{A'}_{\text{In}}^2I(I+1) &=&\mathcal{A}_{\text{In}}^2I(I+1)+\mathcal{A}_{\text{As}}^2J(J+1),
\end{eqnarray}
which gives a $5\%$ increase in the hyperfine constant for indium nuclear spins. However, this does not mean that the new system with only $N$ indium nuclear spins retains all the properties of the initial system.

\bibliography{NSP}
\end{document}